\documentclass[twoside,leqno,twocolumn]{article}

\usepackage[letterpaper]{geometry}
\usepackage[dvipsnames]{xcolor}
\usepackage{amssymb,amsfonts,amsmath} 
\usepackage{enumerate,enumitem,tikz,graphicx,mathrsfs,eucal,verbatim, bbm, derivative}
\usepackage{tkz-graph, tikz-cd}
\usepackage{graphicx}
\usepackage{ltexpprt}
\usepackage{hyperref}

\usepackage{caption}
\usepackage{subcaption}
\usepackage{wrapfig}
\usepackage{graphbox} 
\usepackage{comment}
\usepackage{algorithm}

\usepackage{algpseudocode}
\usepackage{nicefrac}
\usepackage{hhline}

\usepackage{booktabs}
\usepackage{tabularx}
\usepackage{longtable}
\usepackage{multirow}
\usepackage{array}

\usepackage{wasysym}
\usepackage{diagbox}

\usepackage{placeins}

\usepackage[most]{tcolorbox}

\newcommand\Tstrut{\rule{0pt}{2.6ex}}         
\newcommand\Bstrut{\rule[-0.9ex]{0pt}{0pt}}   

\newcommand{\mthd}[1]{\texttt{HyperSteiner}#1}
\newcommand{\abs}[1]{\left\lvert #1 \right\rvert}
\newcommand{\RR}{\mathbb{R}}
\newcommand{\CC}{\mathbb{C}}
\newcommand{\HH}{\mathbb{H}}

\newtheorem{thm}{Theorem}[section]
\newtheorem{prob}[thm]{Problem}

\newtheorem{prop}[thm]{Proposition}

\newtheorem{defn}{Definition}

\begin{document}

\title{\Large HyperSteiner: Computing Heuristic Hyperbolic Steiner Minimal Trees}
\author{Alejandro García-Castellanos \footnotemark[2] \footnotemark[1] \and Aniss Aiman Medbouhi \footnotemark[3] \footnotemark[1] \and Giovanni Luca Marchetti \footnotemark[4] \and Erik J. Bekkers \footnotemark[2] \and Danica Kragic \footnotemark[3]}

\date{}

\maketitle






\begin{abstract} \small\baselineskip=9pt We propose HyperSteiner -- an efficient heuristic algorithm for computing Steiner minimal trees in the hyperbolic space. HyperSteiner extends the Euclidean Smith-Lee-Liebman algorithm, which is grounded in a divide-and-conquer approach involving the Delaunay triangulation. The central idea is rephrasing Steiner tree problems with three terminals as a system of equations in the Klein-Beltrami model. Motivated by the fact that hyperbolic geometry is well-suited for representing hierarchies, we explore applications to hierarchy discovery in data. Results show that HyperSteiner infers more realistic hierarchies than the Minimum Spanning Tree and is more scalable to large datasets than Neighbor Joining.\end{abstract}

\section{Introduction}\label{sec:intro}

\renewcommand{\thefootnote}{\fnsymbol{footnote}}\footnotetext[1]{Equal contribution. E-mails: $<$a.garciacastellanos@uva.nl$>$, $<$medbouhi@kth.se$>$.}\footnotetext[2]{Amsterdam Machine Learning Lab, University of Amsterdam, Netherlands.}\footnotetext[3]{School of Electrical Engineering and Computer Science, KTH Royal Institute of Technology, Sweden.}\footnotetext[4]{School of Engineering Sciences, KTH Royal Institute of Technology, Sweden.}The \emph{Steiner Minimal Tree} (SMT) is a problem in computational geometry consisting of finding a tree of minimal length containing a given set of points as vertices. It is commonly applied to network design of various forms, for example, digital chips and cellular networks. However, finding an SMT is computationally intractable, since the problem is known to be NP-hard in several of its variants \cite{garey1977complexity}. As a consequence, heuristic approaches have been proposed, such as the \emph{Smith-Lee-Liebman (SLL) algorithm} \cite{smith1981n}. Even though the SMT problem is often considered in the Euclidean plane, it is definable in arbitrary geometries. For example, it has been considered for non-Euclidean spaces, such as Riemannian manifolds \cite{ivanov2000steiner, logan2015steiner}. 

\begin{figure}
\centering
\includegraphics[width=.7\linewidth]{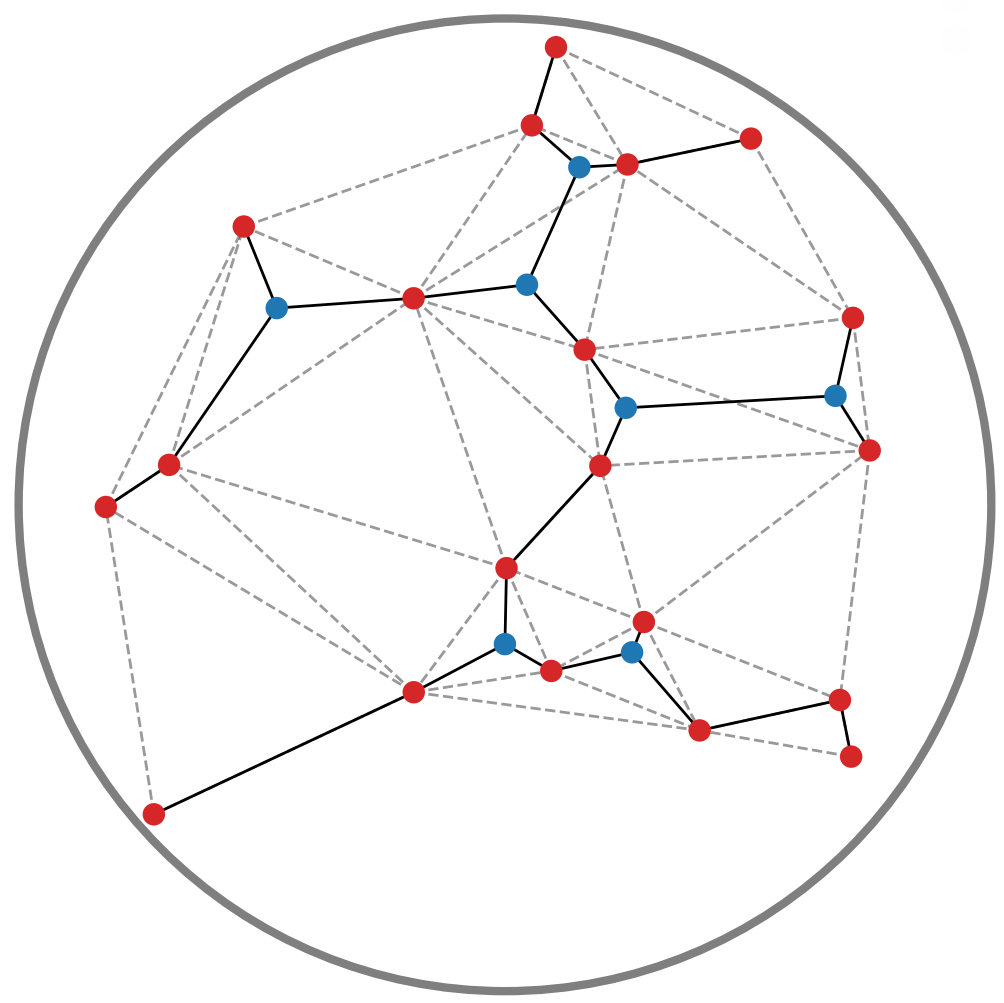}
\caption{A hyperbolic Steiner tree computed via \mthd{}. Red ({\color{Red!80!black}$\newmoon$}) denotes terminals, blue ({\color{NavyBlue}$\newmoon$}) denotes Steiner points, while the dashed line corresponds to the auxiliary hyperbolic Delaunay triangulation.}
\label{fig:firstpage}
\end{figure}

The \emph{Hyperbolic space} is a non-Euclidean geometry that has been extensively deployed to represent tree-like structures. The latter are typical of data organized in a \emph{hierarchical} manner. The hyperbolic space is particularly suitable for representing trees due to its negative curvature, which forces the volume to concentrate towards the boundary, accommodating arbitrary branching \cite{sarkar2011low}. This has motivated the development of machine learning methods for hyperbolic embeddings of hierarchical data \cite{nickel2017poincare, chamberlain2017neural, mathieu2019continuous}. 

Hierarchies emerge in several domains. In natural language, semantics induces a hierarchy where words corresponding to specialized concepts lie towards the leaves of the tree \cite{tifrea2018poincar, nickel2017poincare}. In 
phylogenetics, the branching of biological evolution determines a tree over genetic data \cite{Cieslik2006shortestconnectivity}. Lastly, in the context of games and reinforcement learning, the state space branches when the agent performs an action or when an event occurs in the environment \cite{cetin2022hyperbolic}. In all of these domains, hyperbolic data representations provide a fundamental tool that can be exploited for data analysis and inference.

In this work, we propose \mthd{} -- an efficient heuristic algorithm for addressing the SMT problem in the two-dimensional hyperbolic space (see Figure~\ref{fig:firstpage}). Differently from the Euclidean case, in hyperbolic geometry two dimensions suffice to embed any tree with arbitrary low distortion \cite{sarkar2011low}. Motivated by this, we propose a hyperbolic version of the SLL algorithm. The latter is grounded in a divide-and-conquer approach via constructing a Delaunay triangulation and solving local SMT problems over the Delaunay simplices. \mthd{} computes the hyperbolic Delaunay triangulation via Euclidean power cells and phrases the local SMT problems in an algebraic fashion as a system of polynomial equations, which is then solved numerically. The local solutions are finally aggregated in order to obtain a heuristic SMT. We deploy the \emph{Klein-Beltrami model} of hyperbolic geometry since it enables us to compute efficiently both the Delaunay triangulation and the algebraic expressions required.  

As a main application, we deploy \mthd{} for \emph{hierarchy discovery} in real-life biological data. To this end, we build a heuristic SMT on top of a hyperbolic data representation, and exploit it to infer the age of cells of the \emph{Planaria} species \cite{Plass2018_planaria_SingleCellData}. We show empirically that \mthd{} performs better than the Minimum Spanning Tree in terms of cell age prediction. Moreover, it is significantly more computationally efficient than Neighbor Joining \cite{Saitou1987neighborjoining} -- a popular tree reconstruction method in phylogenetics.

In summary, our contributions are:
\begin{itemize}
\item A version of the SLL algorithm for the hyperbolic space named \mthd{}. 
\item An implementation of \mthd{} via numerical solvers for systems of algebraic equations. 
\item An empirical investigation on synthetic and real data, with applications to hierarchy discovery in biology.
\end{itemize}
Our Python implementation of \mthd{} and the experiments is provided at the following public repository: \url{https://github.com/AGarciaCast/HyperSteiner}.

\section{Related Work}
In this section, we review the relevant literature around Polynomial-Time Approximation Schemes (PTAS) for SMTs, the SMT problem in non-Euclidean geometries, the deployment of hyperbolic geometry in machine learning, and the methods for hierarchy discovery in data.\\  

\paragraph{Polynomial-Time Approximations of SMTs.}
A Polynomial-Time Approximation Schemes (PTAS) for a given problem is a potentially-probabilistic polynomial-time algorithm that solves the problem approximately within arbitrary error margins. Specifically, for each $c > 1$, a PTAS can approximate the solution to the problem within a factor of $1 + 1/c$ in polynomial time. It has been shown that the Euclidean SMT problem admits a PTAS \cite{Arora1998ProofVerificationHardness, Arora1998PTASTSP, Mitchell1999PTASTSP}, with several subsequent improvements \cite{Rao1998AroraImprovement, Borradaile2009AroraImprovement}. The time complexity of these algorithms is exponential in $c$ -- for example, in \cite{Arora1998PTASTSP} it is $\mathcal{O}(|P| ({\log |P|})^{\mathcal{O}(c)})$ for the two-dimensional case, where $|P|$ is the number of terminals. Moreover, these methods are often probabilistic, admitting de-randomized versions that are significantly less efficient \cite{Czumaj1998PTASdeterministicArora}. Overall, all the PTASs depend on an error $c$ that can be controlled, but that affects exponentially the computational complexity. In contrast, the SLL algorithm -- and our hyperbolic adaptation \mthd{} -- is deterministic, and no explicit tight bound on its error is available, since it relies on the Delaunay triangulation as a heuristic. However, as a major advantage, the complexity is $\mathcal{O}(|P| \log |P|)$, independently of the error -- see Section~\ref{sec:compcomp} for details.\\

\paragraph{Non-Euclidean SMTs.}
The SMT problem can be defined in arbitrary metric spaces \cite{cieslik2001steiner} and has been studied for several non-Euclidean metrics. Examples are the Manhattan (i.e., the $l_1$) metric \cite{hanan1966steiner}, in which case the SMT is referred to as \emph{rectilinear}, orientation metrics \cite{yan1997steiner}, and Riemannian manifolds \cite{ivanov2000steiner, logan2015steiner}. The above-mentioned PTAS \cite{Arora1998PTASTSP} has not yet been adapted to the hyperbolic SMT problem. Instead, the SLL algorithm for computing heuristic SMTs has been extended to spherical geometry \cite{dolan1991minimal}, while explicit constructions for full Steiner trees in hyperbolic geometry have been proposed \cite{halverson2005steiner}. Our work contributes to this line of research by extending the SLL algorithm to the hyperbolic space. Since the SLL algorithm involves solving local full SMT problems, the algorithm we propose incorporates techniques closely related to the ones from \cite{halverson2005steiner}. However, to the best of our knowledge, our approach is the first to present a computationally feasible algorithm for heuristic hyperbolic SMTs. Indeed, the algebraic method proposed by \cite{halverson2005steiner} for computing exact hyperbolic SMTs in the upper half-plane model is unable to handle more than 4 terminals in practice.\\

\paragraph{Hyperbolic Machine Learning.} Hyperbolic geometry has gained interest in machine learning due to its effectiveness in handling data with hierarchical structures. Several methods have seen extensions to hyperbolic geometry \cite{ganea2018hyperbolic, peng2021hyperbolic}, especially in the context of generative modeling and representation learning. Hyperbolic analogs have been designed for generative models such as variational auto-encoders \cite{mathieu2019continuous, Nagano2019wrapped}, normalizing flows \cite{bose2020latent}, and diffusion models \cite{huang2022riemannian}. Moreover, manifold learners embedding data in a hyperbolic space have been developed and applied to various domains \cite{nickel2017poincare, chamberlain2017neural, Sala2018_RepresentationTradeoffsHyperbolicEmbeddings, Chami_2021_HoroPCA, Guo_2022_CO-SNE}. Once data is embedded in a hyperbolic latent space, efficient and geometry-driven statistical inference can be performed \cite{Medbouhi2024hHyperDGA}. Our work adheres to this paradigm, since heuristic SMTs of data can be computed in a hyperbolic latent space via \mthd{} and used for applications such as hierarchy discovery.\\

\paragraph{Hierarchy Discovery.}
The problem of hierarchy discovery consists of reconstructing a tree-like structure from data possessing an unknown hierarchy. It has seen applications in evolutionary biology for phylogenetic trees \cite{Cieslik2006shortestconnectivity} and, more recently, in cell lineage reconstructions due to progress in molecular sequencing \cite{Gong20218benchmarkcelllineage}. For example, given an evolutionary model that penalizes some mutations in sequence data -- typically, DNA, RNA, or proteins \cite{Miguel2019EvolutionaryModels} -- one can perform sequence alignment \cite{Needleman1970SequenceAlignment1, Smith1981SequenceAlignment2} in order to estimate the evolutionary distance between data points, from which a phylogenetic tree can be built. Tree reconstruction methods in this spirit are deemed \emph{distance-based}. They include Unweighted Paired Group Mean Arithmetic (UPGMA) \cite{Sokal1958UPGMA} and Neighbor Joining \cite{Saitou1987neighborjoining}. Alternative \emph{character-based} methods are probabilistic and involve comparing all data sequences simultaneously. They include Maximum Parsimony \cite{Fitch1971maximumparsimony}, Maximum Likelihood \cite{Felsenstein1981MaximumLikelihoodPhylogenetic}, and Bayesian methods \cite{Huelsenbeck2001BayesianPhylogenetic}. Reconstructing a phylogenetic tree can be formulated as a Steiner tree problem by minimizing the number of mutations \cite{Foulds1979steinerminimalphylogenetictree} -- a research direction that has been investigated for Euclidean data \cite{Blelloch1997PhylogeneticSteinerTree, Brazil2009PhylogeneticSteinerTree, Weng2011probabilitysteinerphylogenetic, FA2022steinertreelineagecell}. \mthd{} efficiently computes a hyperbolic Steiner tree, which is a way to discover the hierarchy for data represented in a hyperbolic space. The latter aligns with recent hyperbolic embeddings for biological data \cite{Klimovskaia2020SingleCellPoincareMap, Zhou2021HyperbolicGeometryGeneExpression}.

\section{Background}
In this section, we recall the necessary background around Steiner trees and hyperbolic geometry. 

\subsection{Steiner Minimal Trees.}\label{sec:steimin}
\begin{sloppypar}
We begin by introducing the SMT problem in a general metric space. Intuitively, the problem consists in minimizing the total edge length of a tree, some of whose vertices are fixed, while others are allowed to be added in the ambient space. Let $\mathcal{X}$ be a metric space with distance function ${d\colon \mathcal{X} \times \mathcal{X} \to \RR_{\geq 0}}$. 
\end{sloppypar}
 
\begin{prob}\label{prob:mst}
Given a finite set $P \subseteq \mathcal{X}$, find a finite undirected connected graph $\text{\rm SMT}(P)=(V, E)$ embedded in $\mathcal{X}$ such that $P \subseteq V$ and the total edge length, defined as follows, is minimal:
\begin{equation}
L(\text{\rm SMT}(P)) = \sum_{(x,y) \in E} d(x,y).
\end{equation}
\end{prob}
The minimal graph is necessarily a tree, since edges can be removed from cycles without disconnecting the graph. The elements of $P$ are traditionally referred to as \emph{terminals}, while the additional points in $V \setminus P$ are auxiliary ones referred to as \emph{Steiner points}. The number of Steiner points is bounded by $|P| - 2$, and when the bound is reached, we call the Steiner tree \emph{full} (FST). When $|P|=3$, the only Steiner point corresponds to the \emph{Fermat point} of the triangle with vertices $P$. When $|P|=4$, an FST contains two Steiner points, and its topology must coincide with one of the two displayed below (square denotes terminals while star denotes Steiner points). 
\begin{figure}[ht!]
\centering
  \begin{subfigure}[t]{0.2\linewidth}
     \centering
     \includegraphics[width=\linewidth]{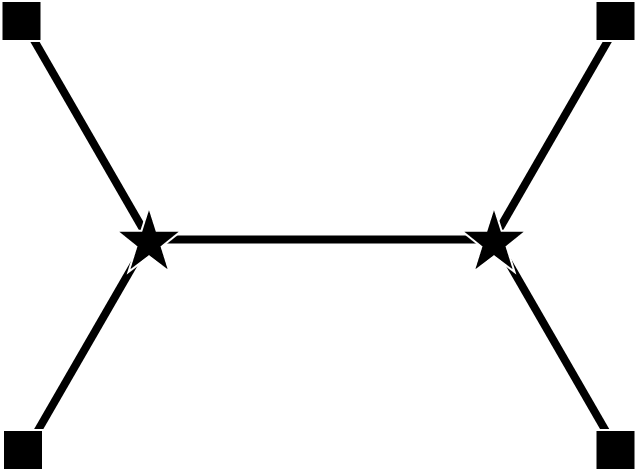}
 \end{subfigure}
 \hspace{4em}
 \begin{subfigure}[t]{0.2\linewidth}
     \centering
     \includegraphics[width=\linewidth]{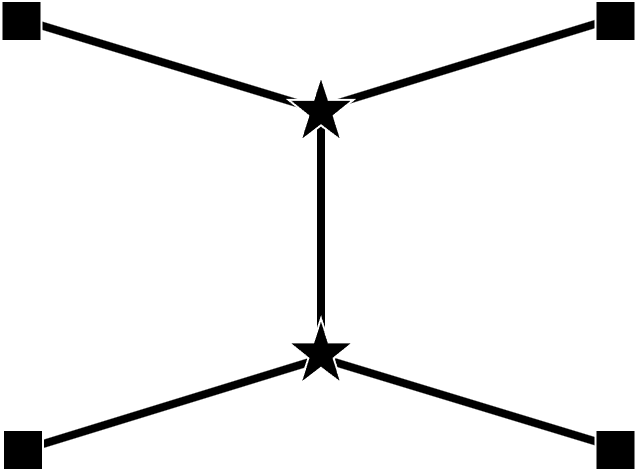}
 \end{subfigure}

\end{figure}

With the condition $P = V$, Problem~\ref{prob:mst} reduces to the \emph{Minimum Spanning Tree} (MST) problem. The latter can be solved efficiently via Kruskal's algorithm \cite{kruskal1956shortest} in $\mathcal{O}(|P|^2 \log |P|)$ time. However, the SMT is instead NP-hard \cite{garey1977complexity}. 

An SMT has total edge length lower than an MST. The following quantity, which will be necessary in the upcoming constructions, compares the two edge lengths.  
\begin{defn}\label{locstdef}
\begin{sloppypar}
The \emph{local Steiner ratio} for a subset ${P\subseteq \mathcal{X}}$ is:    

\begin{equation}\label{eq:stratio}
\rho(P) = \frac{L(\text{\rm SMT}(P))}{L(\text{\rm MST}(P))} \leq 1.
\end{equation}
\end{sloppypar}

\end{defn}

\subsection{The SLL Algorithm.}\label{sec:sll}
We now provide an overview of the Smith-Lee-Liebman (SLL) algorithm \cite{smith1981n} with a minor modification from \cite{zachariasen_concatenation-based_1999}. The overall idea is to compute a heuristic SMT via a divide-and-conquer approach. Specifically, the terminals $P$ are geometrically partitioned into small subsets, so that local FST problems can be solved over such subsets. The subdivision is performed via a \emph{Delaunay triangulation} (DT), whose definition we recall below. 
\begin{defn}
The \emph{Voronoi cell} of $p \in P$ is defined as:
\begin{equation}
V(p) = \{ x\in \mathcal{X} \ | \ \forall q \in P \  d(x,q) \geq d(x,p) \}.
\end{equation}
\begin{sloppypar}
The \emph{Delaunay triangulation} is defined as the collection of simplices with vertices $\sigma \subseteq P$ such that ${\bigcap_{p \in \sigma } V(p) \not = \emptyset}$.    
\end{sloppypar}

\end{defn}
We now focus on $\mathcal{X} = \RR^2$. Specifically, we will describe the SLL algorithm for the Euclidean plane, although it can be adapted to $\RR^3$ as explained by \cite{smith_steiner_2013}.

\begin{sloppypar}
The SSL algorithm leverages upon the MST as an initial approximation of the SMT. The approach is motivated by the conjecture that the local Steiner ratio for any $P\subset \RR^2$  satisfies $\rho(P) \geq \sqrt{3}/2 \approx 0.866$, i.e., ${L(\text{MST}(P)) \leq 2/\sqrt{3} \  L(\text{SMT}(P))}$\footnote{Up to this date, the Gilbert-Pollak Steiner ratio conjecture remains unsolved \cite{ivanov2012steiner}. The best known lower bound for this value is $\sim 0.824$ \cite{news_steiner_ratio}.}. Based on this, the SSL algorithm replaces the edges of the MST connecting subsets of $3$ and $4$ terminals with their corresponding FSTs. Since the MST is contained in the $1$-skeleton of the Delaunay triangulation \cite{preparata2012computational} (i.e. the 1-dimensional Delaunay edges), the SLL algorithm selects the subsets of replaced terminals among the vertices of Delaunay triangles. We summarize the approach in Algorithm~\ref{alg:sll-alg}.     
\end{sloppypar}

\begin{algorithm}[t!]
    \caption{SLL Algorithm}
    \label{alg:sll-alg}
\vspace{2mm}

\textbf{Input}: Terminals $P \subset \mathbb{R}^2$.
\begin{enumerate}
    \item Construct the Delaunay triangulation, $\text{DT}(P)$.
    \item Construct $\text{MST}(P)$ (Kruskal algorithm) and simultaneously build a priority queue as follows:
    \begin{enumerate}[label*=\arabic*.]
        \item Mark all the triangles $\sigma \in \text{DT}(P)$ containing two edges of $\text{MST}(P)$ and admitting an FST. 
        \item Place the FSTs of marked triangles $\sigma$ in a queue $Q$ prioritized on $\rho(\sigma)$ (smaller first).
    \end{enumerate}
    \item Add the FST of the $4$-terminal subsets:
    \begin{enumerate}[label*=\arabic*.]
        \item For each marked triangle $\sigma$, find its adjacent triangles $\sigma'$ such that $\sigma$ and $\sigma'$ contain three edges of the $\text{MST}(P)$.
        \item Compute the FST $\sigma \cup \sigma'$ for each of the two possible topologies and add the minimal one to $Q$. 
    \end{enumerate}
    \item Convert $Q$ to an ordered list and append to it the edges of $\text{MST}(P)$, sorted in non-decreasing order.
    \item Let $T$ be an empty tree. An FST in $Q$ is added to $T$ if it does not create a cycle (greedy concatenation).  
\end{enumerate}
\textbf{Output}: $T$ -- a heuristic SMT of $P$. 
\vspace{2mm}
\end{algorithm}

\subsection{The Klein-Beltrami Model.}
The $n$-dimensional hyperbolic space is, by definition, the unique simply-connected Riemannian manifold with constant curvature equal to $-1$. It admits several isometric models. In this work, we deploy the Klein-Beltrami model since its geodesics are straight lines, which enables to reduce the derivations and computations to the Euclidean case, in particular for computing the hyperbolic Delaunay triangulation. 

The $n$-dimensional Klein-Beltrami model is defined as the Riemannian manifold with ambient space the Euclidean unit ball $\mathbb{K}^{n}= \left\{z \in \RR^n \mid\|z\|_2 <1\right\}$,  equipped with the metric tensor:
\begin{equation}
 g(z) = \left( \frac{-1}{\langle z,z \rangle}I_{n} + \frac{1}{\langle z,z \rangle^2} z\otimes z \right),
 \end{equation}

\begin{sloppypar}
where $\otimes$ denotes the tensor product, $I_n$ denotes the identity matrix, and $\langle \cdot, \cdot \rangle$ denotes the \emph{Lorentzian Inner Product in homogeneous coordinates} i.e., $\langle \cdot, \cdot \rangle$ equals $-1$ plus the Euclidean scalar product. As for any Riemannian manifold, $\mathbb{K}^n$ can be seen as a metric space when equipped with the geodesic distance ${d(x,y) = \inf_\gamma \int_{[0,1]} \sqrt{\gamma'(t)^\dagger g(\gamma(t))\gamma'(t)} \ \textnormal{d}t}$, where ${\gamma\colon [0,1] \rightarrow \mathbb{K}^n}$ is a smooth curve with ${\gamma(0) = x}$, ${\gamma(1)=y}$, and $\gamma'$ denotes the first derivative of $\gamma$. Explicitly, the distance is given by the expression:   
\end{sloppypar}

\begin{equation}\label{eq:KleinDistance}
d(x,y)= \operatorname{arccosh}\left(\frac{-\langle x,y\rangle}{\sqrt{\langle x,x \rangle \langle y,y \rangle}}\right).
\end{equation}

\section{Method}
We now present our method, deemed \mthd{}, for computing heuristic hyperbolic SMTs. Specifically, we propose an adaptation of the SLL algorithm described in Section~\ref{sec:sll} to the Klein-Beltrami model. Our approach is generalizable to other hyperbolic models, as shown in the Appendix (Section~\ref{sec:appother}). In particular, Section~\ref{sec:klein-upper} provides a comparative discussion with the upper half-plane model.

To begin with, the MST can be computed via the Kruskal algorithm in the Klein-Beltrami model since an explicit expression for the distance is available (see Equation~\ref{eq:KleinDistance}). Moreover, similarly to the Euclidean case, the hyperbolic MST is contained in the hyperbolic Delaunay triangulation \cite{sarkar2011low}. Therefore, the challenges of adapting the SLL algorithm are reduced to:
\begin{itemize}
\item Computing the hyperbolic Delaunay triangulation.
\item Computing FSTs for $3$ and $4$ terminals.
\end{itemize}

Below, we describe our proposed solutions to the above points in detail.  

\subsection{Hyperbolic Delaunay Triangulation.}\label{sec:hyperdel}
As anticipated, the Delaunay triangulation is feasible to compute in the Klein-Beltrami model. As shown by \cite{HyperbolicVoronoiDiagramsMadeEasy2010}, the hyperbolic Voronoi cells in the Klein-Beltrami model correspond to weighted Euclidean Voronoi cells, referred to as power cells. We introduce the latter below in a general metric space $\mathcal{X}$.

\begin{defn}\label{def:powercell}
The \emph{power cell} of $p \in P$ with weights $\{ r_p\}_{p \in P} \subseteq \RR_{\geq 0}$ is defined as:
\begin{equation}
R(p) = \{ x\in \mathcal{X} \ | \ \forall q \in P \  d_q(x) \geq d_p(x) \},
\end{equation}
where $d_p(x) = d(x,p)^2 - r_p^2$. 
\end{defn}
For vanishing weights, power cells specialize to Voronoi cells. By leveraging on the non-isometric embedding of the Klein-Beltrami model into the Euclidean space, it is possible to reduce the computation of hyperbolic Voronoi cells to Euclidean power cells with appropriate points and weights. 

\begin{thm}[\cite{HyperbolicVoronoiDiagramsMadeEasy2010}]\label{thm:hypervoronoi}
Given $P \subseteq \mathbb{K}^n$, there exists an explicit set $S \subseteq \RR^n$ and weights $\{ r_s\}_{s \in S}$ such that the hyperbolic Voronoi cells of $P$ correspond to restrictions to $\mathbb{K}^n$ of power cells of $S$. 
\end{thm}

Power cells, together with their intersections, can be computed in the Euclidean space via standard algorithms from computational geometry. This enables us to obtain hyperbolic Voronoi cells via Theorem~\ref{thm:hypervoronoi}. 

\subsection{Fermat Points.}\label{sec:fermatpts}

In this section, we describe how to compute Fermat points in the Klein-Beltrami model. Similarly to the SLL algorithm (Section~\ref{sec:sll}), we will focus on the two-dimensional case $\mathcal{X} = \mathbb{K}^2$. We start with a general result for Riemannian surfaces.

\begin{thm}[\cite{Weng2001SteinerTreesCurvedSurfaces}]\label{thm:steinertreeangle120}
All the angles formed by the edges of an SMT on a Riemannian surface are no less than $2\pi/3$. In particular, all the angles at Steiner points are equal to $2\pi/3$.
\end{thm}

This motivates the introduction of the following curves on Riemannian surfaces.  

\begin{defn}\label{def:isoptic}
Let $\mathcal{X}$ be a complete Riemannian surface and $x,y \in \mathcal{X}$. The \emph{isoptic curve} with an angle $\alpha$, denoted by $C_\alpha(x,y)$, is the locus of the points $s \in \mathcal{X}$ such that some geodesic connecting $s$ with $x$ forms an angle of $\alpha$ in $s$ with some geodesic connecting $s$ with $y$. 
\end{defn}

\begin{figure}[ht!]
\centering
  \begin{subfigure}[t]{0.5\linewidth}
     \centering
     \includegraphics[width=\linewidth]{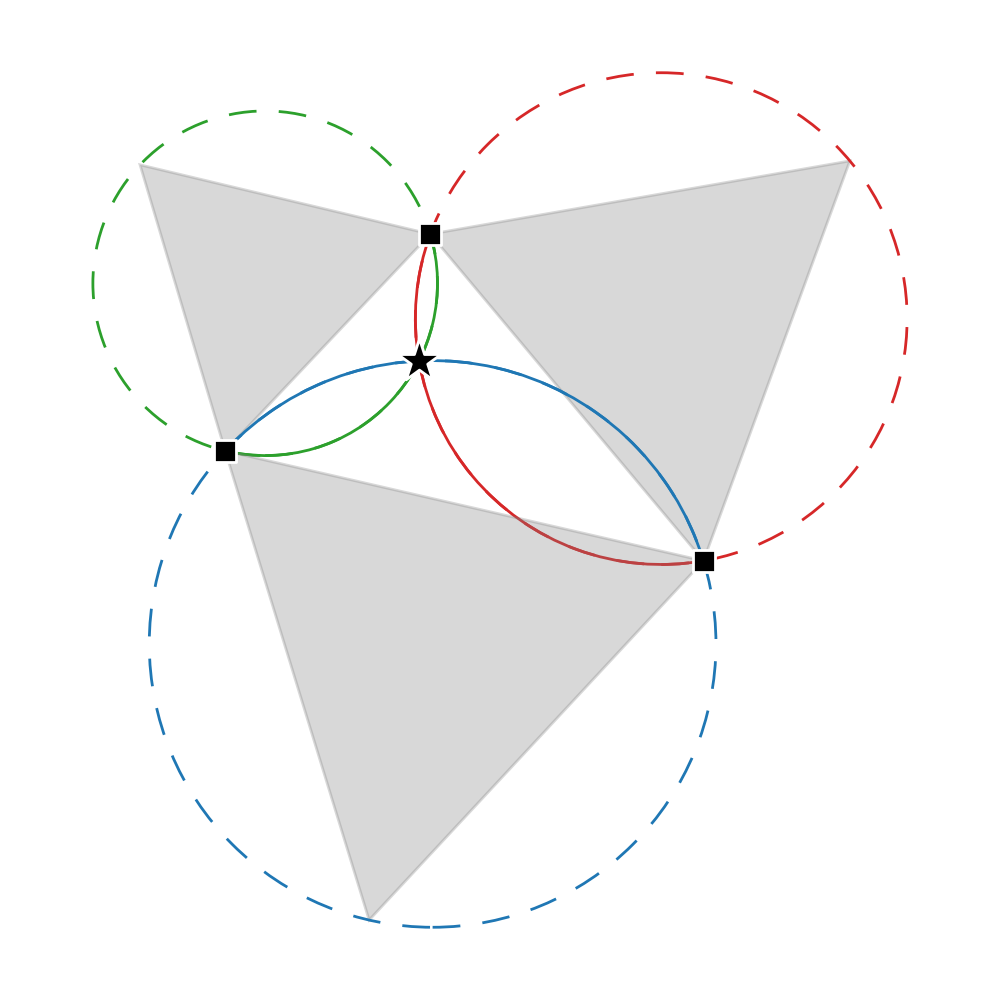}
     \caption{Euclidean, with Torricelli's construction}
     \label{fig:isopticsEuc}
 \end{subfigure}\hfill
 \begin{subfigure}[t]{0.5\linewidth}
     \centering
     \includegraphics[width=\linewidth]{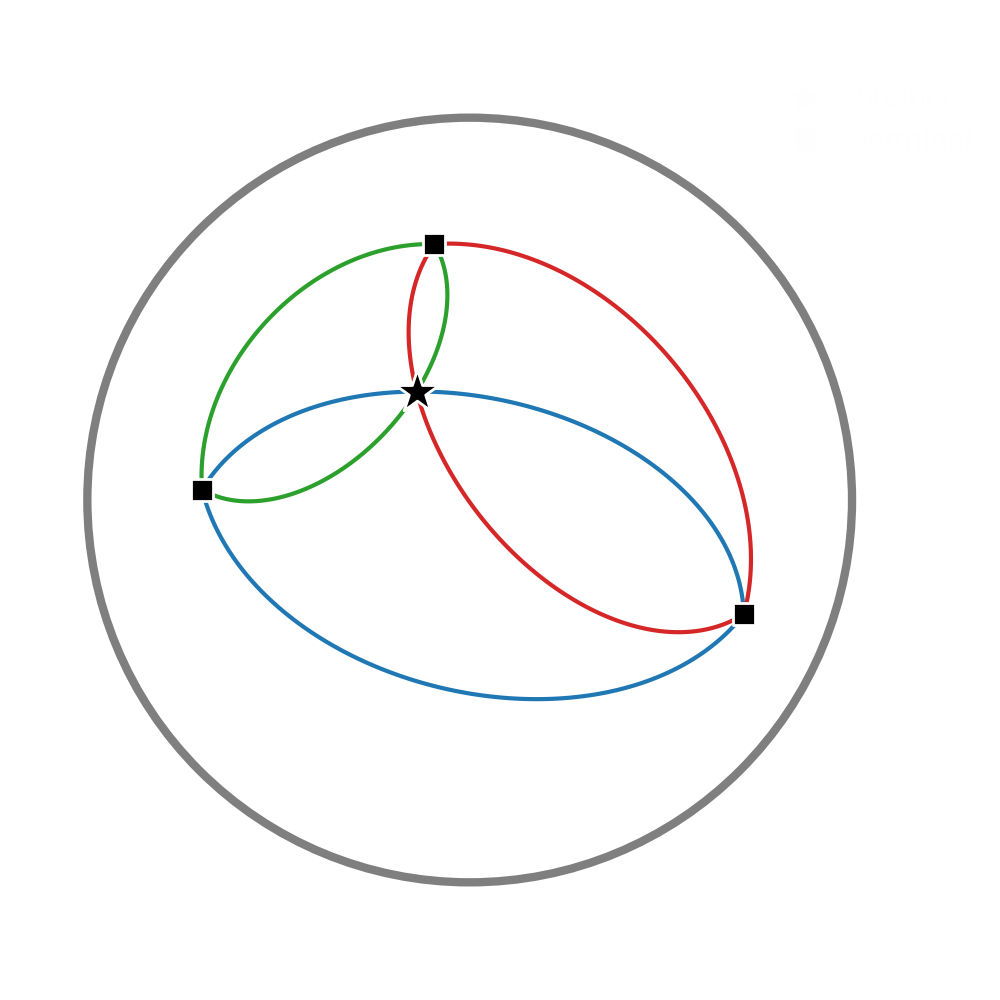}
     \caption{Klein-Beltrami}
     \label{fig:isopticsKlein}
 \end{subfigure}
\caption{Example of isoptic curves for $\alpha=2\pi/3$. Each color represents a different curve. Square ($\,\blacksquare\,$) denotes terminals while star ($\bigstar$) denotes Steiner points.}
\label{fig:isoptics}
\end{figure}

Isoptic curves can be defined more generally with respect to a base curve -- see \cite{geza_isoptic_2020}. In this generality, Definition~\ref{def:isoptic} corresponds to isoptic curves associated with the geodesic connecting $x$ to $y$. When $\mathcal{X}=\RR^2$, the isoptic curves correspond to circular arcs. As a consequence of Proposition~\ref{thm:steinertreeangle120}, Fermat points can be characterized as an intersection of three isoptic curves. Specifically, suppose that $P = \{x, y, z \} \subseteq \mathcal{X}$ are three terminals on a complete Riemannian surface. Then the Fermat point $s$ satisfies: 
\begin{equation}\label{eq:inters}
\{ s\} = C_{\frac{2 \pi}{3}}(x,y) \cap C_{\frac{2 \pi}{3}}(y,z) \cap C_{\frac{2 \pi}{3}}(z,x).
\end{equation}
Figure~\ref{fig:isopticsKlein} provides a visualization of the intersection above. This generalizes the well-known Torricelli method for obtaining Fermat points in the Euclidean plane as an intersection of three circular arcs \cite{promel_steiner_2002} (see Figure~\ref{fig:isopticsEuc}).

We now derive an explicit expression for isoptic curves in the Klein-Beltrami model. Proposition~\ref{prop:isoptic} generalizes previous work on isoptic curves in hyperbolic geometry \cite{csima_isoptic_2012, csima_isoptic_2013, geza_isoptic_2020}.

\begin{prop}\label{prop:isoptic}

Given two points $x, y \in \mathcal{X}=\mathbb{K}^2$ and an angle $0<\alpha<\pi$, the isoptic curve $C_\alpha(x,y)$ in the Klein-Beltrami model is given by $\varphi_{x,y, \alpha}(s) = 0$, where: 
 \begin{align*}
&  \varphi_{x,y, \alpha}(s) =  \langle x, s\rangle \langle y,s \rangle - \langle x,y \rangle \langle s,s \rangle - \\
&- \cos(\alpha)\sqrt{\left(\langle x,s \rangle^2 - \langle x,x \rangle\langle s,s \rangle \right) \left(\langle y,s \rangle^2 - \langle y,y \rangle\langle s,s \rangle \right)}.
\end{align*}
\end{prop}
We provide a proof in the Appendix (Section~\ref{sec:appisop}). Since only two out of three isoptic curves are enough to determine the intersection from Equation~\ref{eq:inters}, finding the Fermat point $s$ of $P = \{x, y, z \} \subseteq \mathbb{K}^2$ is equivalent to solving, for example, the system of equations:
\begin{equation}\label{eq:system}
\begin{cases}
\varphi_{x,y,\frac{2\pi}{3}}(s) = 0 \\
\varphi_{y,z,\frac{2\pi}{3}}(s) = 0.
\end{cases}
\end{equation}
The above system can be solved via numerical approximate methods such as Powell's hybrid algorithm \cite{Powell1970NumericalMethodsNonlinearAlgebraicEquations} -- a high-dimensional extension of Newton's method for root finding. 

\begin{figure}[ht!]
    \centering
    \includegraphics[width=0.55\linewidth]{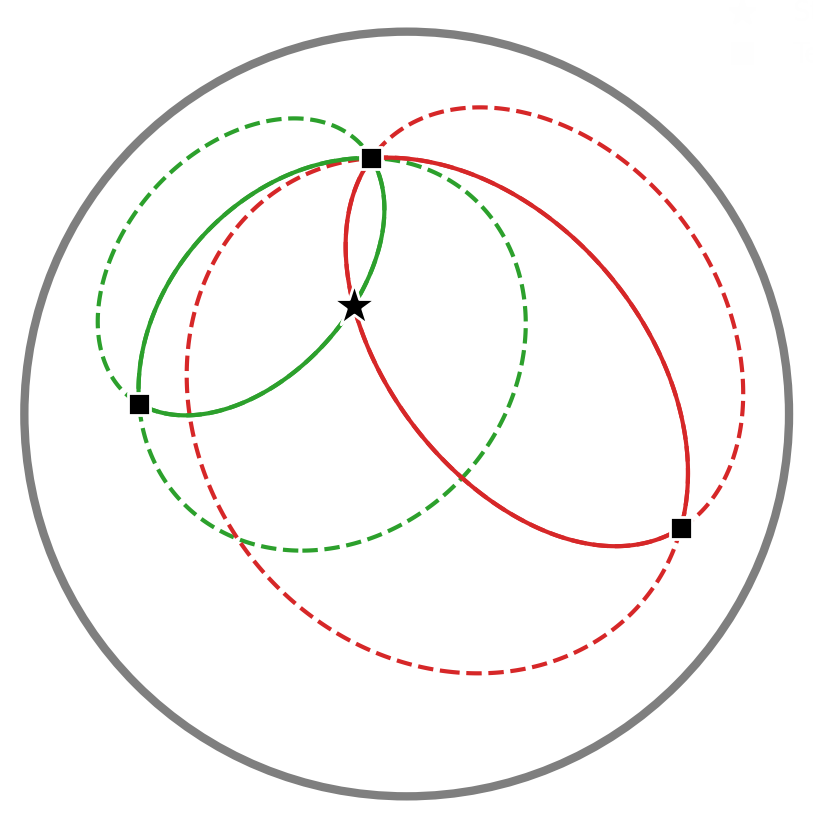}
    \caption{Example of the algebraic curves $\psi$. The solid line corresponds to the isoptic curve for $\alpha=2\pi/3$, and the dashed line to the one for the complementary angle.}
    \label{fig:polySystem}
\end{figure}

A drawback of isoptic curves in the form of Equation~\ref{eq:system} is that the expression is not \emph{algebraic} since it contains square roots. Algebraic formulations are desirable since systems of polynomial equations admit specialized solvers -- for example, the Polynomial Homotopy Continuation (PHC) method from numerical algebraic geometry \cite{chen2015homotopy}. Differently from Newton-based methods, these solvers have certifiable error and convergence guarantees, at the cost of a higher compute. In order to obtain an algebraic expression, the square root from Equation~\ref{eq:system} can be removed via squaring. This, however, introduces additional points in the locus -- see Figure~\ref{fig:polySystem} for a visualization. Specifically, since $\operatorname{cos}(\alpha)^2 = \operatorname{cos}\left(\pi-\alpha\right)^2$, squaring leads to an expression for the union of isoptic curves:
\begin{equation}
C_{\alpha}(x,y) \cup C_{\pi - \alpha}(x,y).
\end{equation}
We summarize the result in the following statement. 
\begin{prop}\label{prop:quadricklein}
\begin{sloppypar}
Given two points ${x,y \in \mathcal{X}=\mathbb{K}^2}$ and an angle ${0<\alpha<\pi}$, the union of isoptic curves ${C_{\alpha}(x,y) \cup C_{\pi - \alpha}(x,y)}$ in the Klein-Beltrami model can be expressed as $\psi_{x,y, \alpha}(s) = 0$, where $\psi_{x,y, \alpha}$ is a quartic polynomial.  
\end{sloppypar}
\end{prop}
We provide the exact expression for $\psi_{x,y,\alpha}$ in the Appendix (Section~\ref{sec:appisop}). 

\subsection{4-Terminal Case.}\label{sec:4terms}

As explained by \cite{halverson2005steiner}, the case $|P|=4$ can be addressed by iteratively computing Fermat points. Specifically, consider a complete Riemannian surface $\mathcal{X}$ and $P = \{ w,x,y,z\} \subseteq \mathcal{X}$. As mentioned in Section~\ref{sec:steimin}, there are two possible full topologies for the FST. Let $a_1$ be a point on $C_{2\pi/3}(w,x)$, and $b_1$ be the Fermat point of $\{a_1, y,z\}$. Define recursively two sequences as follows: $a_{i+1}$ is the Fermat point of $\{w,x,b_i \}$, while $b_{i+1}$ is the Fermat point of $\{y,z,a_{i+1}\}$. Analogously, let $c_1$ be a point on $C_{2\pi/3}(w,y)$, and $d_1$ be the Fermat point of $\{c_1, x,z\}$. Define recursively $c_{i+1}$ as the Fermat point of $\{w,y,d_i \}$, and $d_{i+1}$ as the Fermat point of $\{x,z,c_{i+1}\}$. 

\begin{thm}[\cite{halverson2005steiner}]\label{thm:4points}
The pairs of sequences $\left( \{a_n\}, \{b_n\} \right)$ and $\left( \{c_n\}, \{d_n\} \right)$ converge, respectively, to the two Steiner points of $P$ for each of the two topologies.
\end{thm}

The result above, together with the techniques from the previous section, enables us to compute FSTs with $4$ terminals in $\mathbb{K}^2$. This concludes the description of \mthd.

\subsection{Approximation Error and Computational Complexity.}\label{sec:compcomp}
In this section, we briefly discuss the approximation error of the heuristic SMT found by \mthd{}, as well as the computational complexity of the latter. To begin with, the local Steiner ratio (Definition \ref{locstdef}) in the hyperbolic space satisfies $\rho(P) \geq \frac{1}{2}$ for all $P$ \cite{innami_steiner_2006}. Let $\widehat{\text{SMT}}(P)$ be the tree found by \mthd{} for a given set $P$ of terminals. Since the worst case scenario occurs when no Steiner points are added to the MST by \mthd{}, we obtain  
\begin{equation}
    L(\widehat{\text{SMT}}(P)) \leq L(\text{MST}(P)) \leq 2 \  L(\text{SMT}(P)).
\end{equation}
This provides a bound on the approximation error, in terms of the total edge length, of the heuristic SMT found by \mthd{} w.r.t. the true SMT. 

It is important to note that in hyperbolic geometry, computing the SMT yields the greatest benefit, as it offers the highest maximum percentage of improvement over the MST among any complete surface without a boundary \cite{innami_steiner_2006}.

\begin{sloppypar}
We now analyze the asymptotic computational complexity of \mthd{} w.r.t. the number of terminals. First, the power cells necessary for the hyperbolic Delaunay triangulation can be computed in $\mathcal{O}(|P|\log |P|)$ time by, for example, lifting the points to a hyperboloid and constructing a convex hull \cite{edelsbrunner1985voronoi}. Second, in order to compute the MST, recall that, as mentioned in Section~\ref{sec:steimin}, the Kruskal algorithm has complexity $\mathcal{O}(|P|^2 \log |P|)$. However, since the hyperbolic MST is a subgraph of the hyperbolic Delaunay triangulation \cite{sarkar2011low}, it is possible to restrict the Kruskal algorithm to the $\mathcal{O}(|P|)$ edges of the triangulation, lowering the complexity of computing the MST to $\mathcal{O}(|P| \log |P|)$. Lastly, note that solving the FST problem for ${|P| \in \{3,4 \}}$ has constant complexity w.r.t. $|P|$. The number of FSTs considered is bounded by the number of Delaunay triangles, which in turn is $\mathcal{O}(|P|)$. In conclusion, the complexity of \mthd{} is  $\mathcal{O}(|P| \log |P|)$, matching other SLL-based methods \cite{halverson2005steiner}. In order to grasp the scalability in practice, we provide an empirical investigation of runtimes in the forthcoming experimental section.
\end{sloppypar}

\begin{figure*}[t!]
\centering
  \begin{subfigure}[t]{0.3\linewidth}
     \centering
     \includegraphics[width=\linewidth]{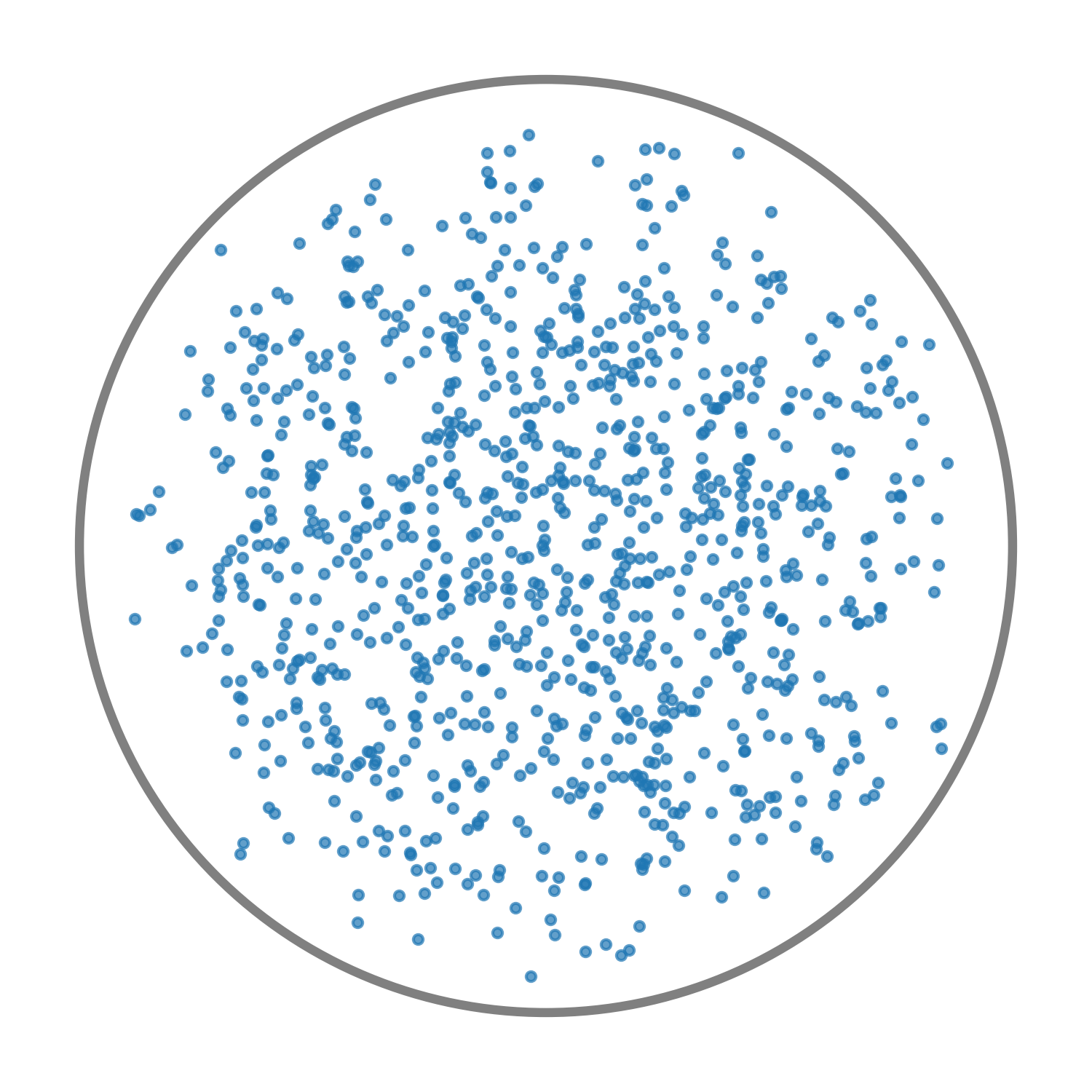}
     \caption*{$\mathcal{G}(0, 0.5 )$}
     \label{fig:samplesTable1}
 \end{subfigure}
 \hfill
 \begin{subfigure}[t]{0.3\linewidth}
     \centering
     \includegraphics[width=\linewidth]{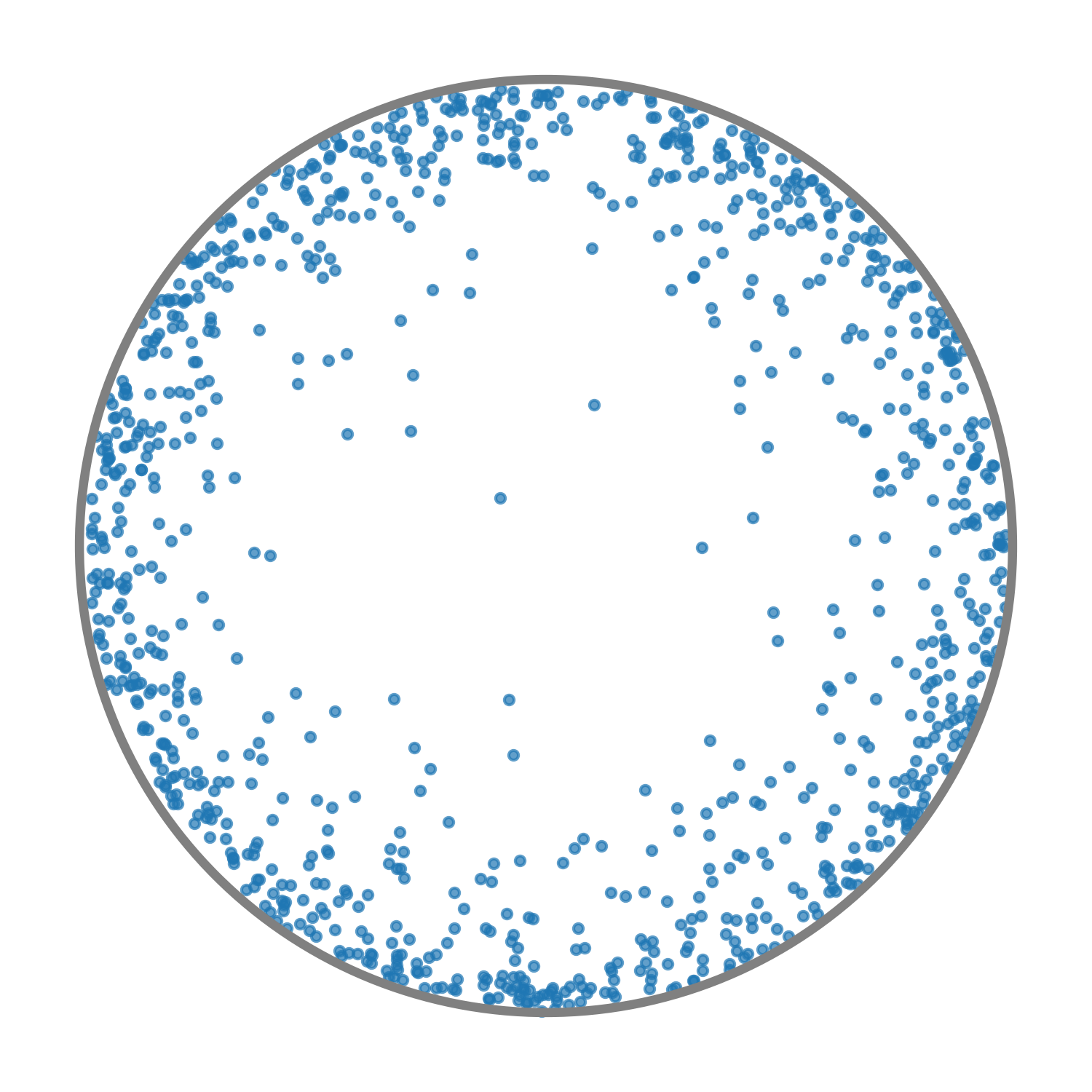}
     \caption*{$\mathcal{G}(\mu_{15, k}(0.9), 0.5)$}
     \label{fig:samplesTable2}
 \end{subfigure}
\hfill
 \begin{subfigure}[t]{0.3\linewidth}
     \centering
     \includegraphics[width=\linewidth]{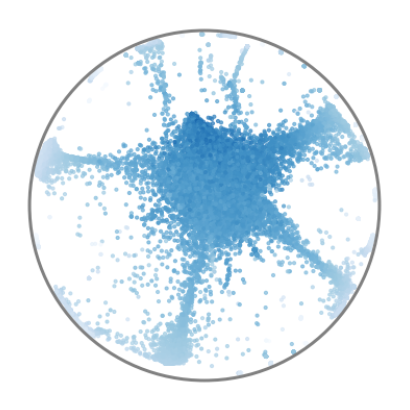}
     \caption*{Planaria}
     \label{fig:samplesTable3}
 \end{subfigure}

\caption{Illustration of the datasets considered. Left and center: $|P|=1000$ random samples from the synthetic datasets. Right: $|P| = 20000$ samples from the real-life dataset, where less saturated color represents older cells. }
\label{fig:samplesTables}
\end{figure*}

\section{Experiments}
In this section, we empirically investigate \mthd{} on both synthetic and real-world data. The goal is twofold. First, to analyze the performance and scalability of the algorithm for different choices of parameters. Second, to showcase an application to hierarchy discovery on a real-life dataset, with comparisons to traditional methods. We implement \mthd{} in Python. The code, including the experiments, is provided in the previously-mentioned public repository. All the experiments were run on Google Cloud via \textsc{n2d} CPUs, with one core and $128$~GB of RAM.

\subsection{Evaluation on Synthetic Data.}
In the first set of experiments on synthetic data, we investigate how the parameters involved in \mthd{} affect its performance and computational complexity. To this end, we generate synthetic datasets by sampling from distributions over $\mathbb{K}^2$. The distributions are constructed from a \emph{pseudo-hyperbolic Gaussian} $\mathcal{G}(\mu, \sigma)$ \cite{Nagano2019wrapped}. The latter is obtained by push-forwarding via the exponential map of the hyperbolic space a Euclidean Gaussian $\mathcal{N}(0, \sigma)$ defined over the tangent space at $\mu \in \mathbb{K}^2$. Based on this, we consider the following two distributions: 
\begin{itemize}
\item A pseudo-hyperbolic Gaussian centered at $\mu = 0$.
\begin{sloppypar}
\item A uniform mixture of pseudo-hyperbolic Gaussians centered at the vertices of a regular ${d\text{-gon}}$ ${\mu_{d,k}(t) = t e^{2 i\pi k /d}}$, with ${k \in \{1, \cdots, d\}}$, for given ${0 < t < 1}$ and $d \in \mathbb{N}$.  
\end{sloppypar}
\end{itemize}

\begin{figure}[!t]
    \centering
    \includegraphics[width=\linewidth]{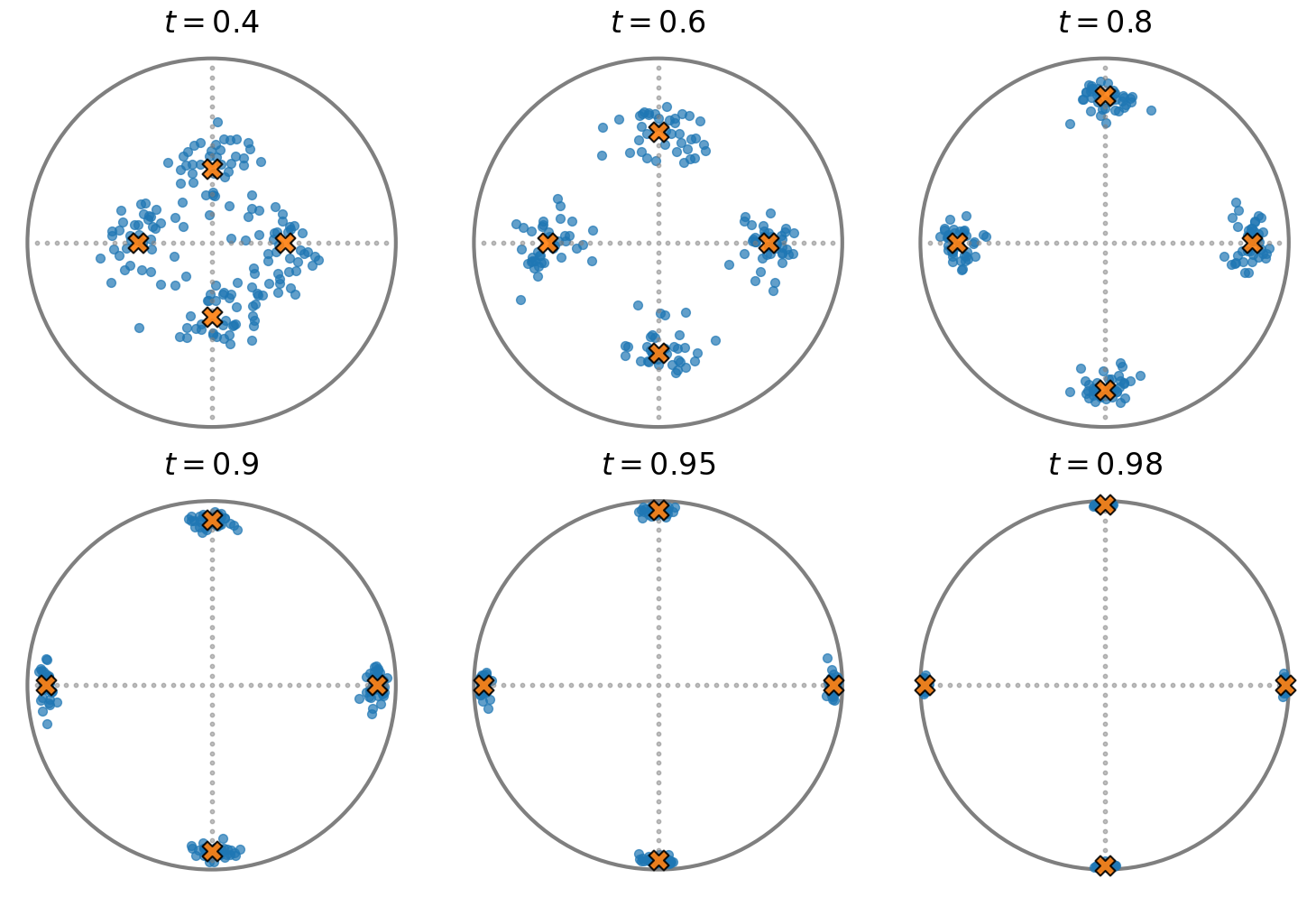}
    \caption{Convergence dataset for $d=4$. Orange crosses represent the means of $\mathcal{G}(\mu_{4, k}(t), 0.15)$.}
    \label{fig:convergence}
\end{figure}

Samples from the centered pseudo-hyperbolic Gaussian can be visualized in Figure \ref{fig:samplesTables} (left). The uniform mixture of pseudo-hyperbolic Gaussians is illustrated in Figure \ref{fig:samplesTables} (center) for a fixed $t$, and in Figure \ref{fig:convergence} for a varying $t$ with four Gaussians. In the scalability analysis experiment with a uniform mixture of pseudo-hyperbolic Gaussians, the distribution, as depicted in Figure \ref{fig:samplesTables} (center), is concentrated towards the ideal boundary of the Klein-Beltrami model, where the volumes increase exponentially. This leads to a scenario where the effects of hyperbolic geometry are more evident. Furthermore, several hyperbolic embeddings from the literature tend to either concentrate the data around the center \cite{Klimovskaia2020SingleCellPoincareMap, nickel_learning_2018} or towards the boundary \cite{nickel2017poincare}.  

\begin{table*}[ht!]
\centering
\resizebox{0.8\textwidth}{!}{
\begin{tabular}{lcccccccccc}
\toprule
\multicolumn{3}{c}{} & \multicolumn{8}{c}{Klein-Beltrami}  \\
\cmidrule(lr){4-11} 
\multicolumn{1}{c}{} & \multicolumn{2}{c}{Euclidean} & \multicolumn{2}{c}{3+Simple} & \multicolumn{2}{c}{3+Precise} & \multicolumn{2}{c}{4+Simple} & \multicolumn{2}{c}{4+Precise} \\
\cmidrule(lr){2-3} \cmidrule(lr){4-5} \cmidrule(lr){6-7} \cmidrule(lr){8-9} \cmidrule(lr){10-11}  
$|P|$ &  RED &      CPU &   RED &      CPU &   RED &      CPU &  RED &      CPU &   RED &       CPU \\
\midrule
50   &  $2.47 \pm 0.55$ &   $0.03$ &    $2.25 \pm 0.61$ &   $0.05$ &   $2.31 \pm 0.61$ &   $0.18$ &   $2.48 \pm 0.61$ &   $0.21$ &  $2.55 \pm 0.60$ &    $1.29$ \\
100   &    $2.46 \pm 0.40$ &   $0.06$ &    $2.17 \pm 0.42$ &   $0.08$ &   $2.21 \pm 0.43$ &   $0.30$  &  $2.39 \pm 0.44$ &   $0.45$ &   $2.45 \pm 0.45$ &    $2.49$ \\
500   &   $2.60 \pm 0.22$ &   $0.37$ &   $2.27 \pm 0.21$ &   $0.42$ &  $2.33 \pm 0.20$ &   $1.71$  &  $2.51 \pm 0.20$ &   $2.28$ &   $2.58 \pm 0.20$ &   $13.07$ \\
1,000  &    $2.62 \pm 0.17$ &   $0.83$ &   $2.32 \pm 0.16$ &   $1.08$ &    $2.37 \pm 0.16$ &   $3.56$ &   $2.56 \pm 0.17$ &   $5.00$ &   $2.62 \pm 0.17$ &   $26.80$ \\
5,000  &  $2.70 \pm 0.07$ &   $9.00$ &   $2.36 \pm 0.07$ &   $7.86$ &   $2.43 \pm 0.07$ &  $19.04$ &   $2.61 \pm 0.07$ &  $29.62$ &   $2.68 \pm 0.07$ &  $137.45$ \\
10,000 &    $2.73 \pm 0.06$ &  $29.84$ &    $2.39 \pm 0.06$ &  $23.15$ &   $2.45 \pm 0.06$ &  $45.52$ &   $2.64 \pm 0.06$ &  $74.05$ &   $2.70 \pm 0.06$ &  $291.53$ \\
\midrule
 Avg.  & $2.60 \pm 0.32$ &   & $2.29 \pm 0.33$ &    & $2.35 \pm 0.33$ &   & $2.53 \pm 0.34$ & & $2.60 \pm 0.34$ & \\
\bottomrule
\end{tabular}
}
    \caption{Scalability analysis of centered Gaussians ($\mathcal{N}(0, 0.5)$ for Euclidean and $\mathcal{G}(0, 0.5)$ for Hyperbolic). RED: reduction over MST (\%). CPU: total CPU time (sec.). For the Euclidean case, we use our own Python implementation of the original SLL algorithm from \cite{smith1981n}.}
\label{tab:scalability_analysisCentered}
\end{table*}

\begin{table*}[ht!]
\centering
\resizebox{0.7\textwidth}{!}{
\begin{tabular}{lcccccccccccccc}
\toprule
\multicolumn{1}{c}{} & \multicolumn{2}{c}{3+Simple} & \multicolumn{2}{c}{3+Precise} & \multicolumn{2}{c}{4+Simple} & \multicolumn{2}{c}{4+Precise} \\
 \cmidrule(lr){2-3} 
 \cmidrule(lr){4-5} 
 \cmidrule(lr){6-7} 
 \cmidrule(lr){8-9} 
 $|P|$  & RED & CPU  & RED &  CPU  & RED & CPU  & RED & CPU \\
\midrule
50    &   $2.61 \pm 0.67$ &   $0.03$ &     $2.75 \pm 0.67$ &   $0.21$ &   $3.00 \pm 0.73$ &   $0.21$ &    $3.18 \pm 0.72$ &    $2.47$ \\
100   &    $2.56 \pm 0.49$ &   $0.07$ &     $2.68 \pm 0.48$ &   $0.41$ &     $2.85 \pm 0.51$ &   $0.46$ &   $2.99 \pm 0.49$ &    $4.21$ \\
500   &    $2.30 \pm 0.18$ &   $0.42$ &   $2.39 \pm 0.19$ &   $2.05$ &     $2.54 \pm 0.20$ &   $2.36$ &     $2.63 \pm 0.21$ &   $18.22$ \\
1,000  &    $2.29 \pm 0.16$ &   $1.06$ &    $2.37 \pm 0.16$ &   $4.36$ &  $2.53 \pm 0.16$ &      $31.49$ &  $2.61 \pm 0.15$ &   $35.16$ \\
5,000  &   $2.34 \pm 0.07$ &   $7.80$ &   $2.39 \pm 0.08$ &  $24.38$ &    $2.57 \pm 0.07$ &  $30.45$  &  $2.62 \pm 0.07$ &  $172.60$ \\
10,000 &    $2.36 \pm 0.05$ &  $22.66$ &   $2.39 \pm 0.04$ &  $56.83$ &    $2.59 \pm 0.05$ &  $71.84$ &    $2.63 \pm 0.05$ &  $345.08$ \\
\midrule
 Avg.  & $2.41 \pm 0.37$ &   & $2.48 \pm 0.38$ &    & $2.68 \pm 0.42$ &   & $2.78 \pm 0.43$ & \\
\bottomrule
\end{tabular}
}
\caption{Scalability analysis of mixture of $\mathcal{G}(\mu_{15, k}(0.9), 0.5)$, $k \in \{1, \cdots, 15\}$. RED: reduction over MST (\%). CPU: total CPU time (sec.). }
\label{tab:scalability_analysisIdeal}
\end{table*}

In the experiments, we focus on two parameters of \mthd{}: the FST number of terminals and the solvers. First, we vary the number of terminals for the exact FST solutions, which can be either $3$ or $4$. In the case of $3$ terminals, Step~3.2 of the SLL algorithm (Algorithm \ref{alg:sll-alg}) is skipped, while for $4$ terminals, the step is addressed via the methods from Section~\ref{sec:4terms}. Second, as discussed in Section~\ref{sec:fermatpts}, we consider two solvers for computing Fermat points. These solvers are the Newton-based method for systems of differentiable equations \cite{Powell1970NumericalMethodsNonlinearAlgebraicEquations}, and the Polynomial Homotopy Continuation (PHC) method for systems of polynomial equations \cite{chen2015homotopy}. We initialize the Newton-based method with two of the considered terminals as starting guesses. This configuration is referred to as the \emph{Simple} setup. It is fast, but lacks convergence guarantees and error bounds. Conversely, we propose an alternative configuration called \emph{Precise} based on the PHC solver. Since relying solely on the latter is computationally expensive, we initially attempt to solve Equation~\ref{eq:system} via the Newton-based method, and if neither of the two solutions corresponds to Steiner points, we employ the PHC solver for the algebraic system of equations from Proposition~\ref{prop:quadricklein}. The \emph{Precise} configuration is more computationally expensive than the \emph{Simple} one, but is able to find accurate solutions, leading to more optimal SMTs.

In order to evaluate the behavior of \mthd{}, we report in Tables \ref{tab:scalability_analysisCentered} to \ref{tab:convergence_analysis}, for the different data distributions mentioned earlier and for different parameters, the percentage of Steiner ratio reduction over the MST (RED), given by $(1 -\rho(P)) \cdot 100$, where $\rho$ is defined in Equation~\ref{eq:stratio}. Moreover, for the scalability analysis experiments, we report in Tables \ref{tab:scalability_analysisCentered} and \ref{tab:scalability_analysisIdeal} the CPU time in seconds. Both the measures are averaged over $100$ random samplings.

\subsubsection{Discussion.}\label{sec:scalability}

From Table~\ref{tab:scalability_analysisCentered}, it is evident that employing the PHC solver and considering $4$-terminal FSTs yields a Steiner ratio reduction similar to that achieved by SLL in the Euclidean space. In the latter, we sample from $\mathcal{N}(0, 0.5)$. However, this setup is more computationally expensive compared to other configurations. Conversely, the Simple version exhibits CPU times comparable to the Euclidean case, albeit at the expense of a worse reduction. When the Simple setup computes $4$-terminal FSTs, it manages to achieve performance akin to its Precise counterpart for all dataset sizes, while still maintaining a lower computational cost.

Additionally, Table~\ref{tab:scalability_analysisIdeal} illustrates that when the terminals are concentrated towards the boundary of $\mathbb{K}^2$, the results obtained are comparable to those reported by \cite{dolan1991minimal}. Notably, the Newton-based method is more prone to failure in such cases, leading to a higher reliance on the PHC solver in the Precise setups. This suggests that a more detailed analysis of the initial guesses used in the Newton-based solver could prove beneficial for further improving performance.

It is worth mentioning that in Tables \ref{tab:scalability_analysisCentered} and \ref{tab:scalability_analysisIdeal}, the improvement of \mthd{} over MST ranges approximately from 2 to 3\%. These results align with previous works studying heuristic SMTs in other geometries. Both in the Euclidean case \cite{smith1981n} and the spherical one \cite{dolan1991minimal}, the improvements obtained fall in the range of 2 to 4\% in terms of RED. We believe that low gains are expected, especially for distributions formed by a highly dense one-connected component, as is the case for Tables \ref{tab:scalability_analysisCentered} and \ref{tab:scalability_analysisIdeal}. The reason for this is the local Euclidean characterization of smooth manifolds, such as the Hyperbolic space.

\begin{table}[ht!]
\centering
\resizebox{\linewidth}{!}{
\begin{tabular}{lcccc}
\toprule
{} & \multicolumn{2}{c}{$d=3$} & \multicolumn{2}{c}{$d=4$} \Bstrut\\
 \cmidrule(lr){2-3}  \cmidrule(lr){4-5} 
\multirow{2}{*}{\backslashbox{$t$}{\strut $|P| / d$}}&           1 &          40  &           1  &          30  \Tstrut\\
 &  &  & & \Bstrut\\
\hline
0.40 &  $13.48 \pm 0.33$ &  $2.52 \pm 0.39$ &   $9.64 \pm 0.27$ &  $2.57 \pm 0.36$ \Tstrut\\
0.60 &  $14.59 \pm 0.19$ &  $2.33 \pm 0.51$ &  $11.55 \pm 0.17$ &  $2.27 \pm 0.37$ \\
0.80 &  $16.31 \pm 0.11$ &  $2.44 \pm 1.27$ &  $14.95 \pm 0.11$ &  $1.89 \pm 0.35$ \\
0.90 &  $17.76 \pm 0.07$ &  $3.16 \pm 2.42$ &  $18.02 \pm 0.08$ &  $1.81 \pm 0.95$ \\
0.95 &  $18.91 \pm 0.05$ &  $4.47 \pm 3.56$ &  $20.54 \pm 0.06$ &  $2.18 \pm 1.93$ \\
0.98 &  $20.04 \pm 0.04$ &  $6.63 \pm 4.56$ &  $23.04 \pm 0.05$ &  $4.02 \pm 4.10$ \\
\bottomrule
\end{tabular}
}
\caption{Convergence analysis for a mixture of $\mathcal{G}(\mu_{d, k}(t), 0.15)$, $k \in \{1, \cdots, d\}$ for $d=3$, $4$ and with an increasing $t$. Values correspond to reduction over MST (\%).  }
\label{tab:convergence_analysis}
\end{table}

In Table~\ref{tab:convergence_analysis}, we consider samples from $\mathcal{G}(\mu_{d, k}(t), 0.15)$ for $d=3$, $4$ with an increasing $t$ 
-- see Figure~\ref{fig:convergence} for an illustration. This is motivated by \cite{innami_steiner_2006}, where it is shown that the infimum of the hyperbolic local Steiner ratio is reached when terminals lie on the vertices of a regular $d$-gon converging to the boundary. This trend is evident in Table~\ref{tab:convergence_analysis}, where the Steiner reduction increases as $t$ tends to $1$. The reduction values for $t \sim 1$ are significantly higher than the ones from Tables \ref{tab:scalability_analysisCentered} and \ref{tab:scalability_analysisIdeal}, suggesting the full advantage of the theoretical results presented in Section~\ref{sec:compcomp} is gained in the presence of clusters close to the boundary. These configurations naturally arise when using techniques such as Hyperbolic Contrastive Learning \cite{yue2023hyperbolic} or Hyperbolic Entailment Cones \cite{ganea2018hyperbolicCone}.

\subsection{Application to real-life data.}

In this section, we explore applications of \mthd{} to hierarchy discovery on real-life data embedded in a hyperbolic space. To this end, we consider \emph{Planaria} data \cite{Plass2018_planaria_SingleCellData} -- a complex and large-scale dataset commonly deployed in phylogenetics. It consists of the RNA-sequencing of 21612 cells of the Planaria flatworm. The data is intrinsically hierarchical, since the cells differentiate progressively from stem cells to more specialized ones, such as epidermal or pharynx cells. The authors provide the pre-processed data with the first 50 principal components obtained via Principal Component Analysis. We use the \emph{Poincaré map} \cite{Klimovskaia2020SingleCellPoincareMap} -- an embedding specifically developed for single-cell data -- to obtain a two-dimensional hyperbolic representation. A visualization of this hyperbolic embedding in the Klein-Beltrami model is given in Figure \ref{fig:samplesTables} (right).

Given the hyperbolic embedding, the goal is to discover the data hierarchy by constructing a tree. The ground-truth hierarchy is unknown, but the age of the cells can be measured via a state-of-the-art \emph{pseudotime estimation} method based on the Partition-Based Graph Abstraction (PAGA) algorithm \cite{Wolf2019_PAGA_graph_abstraction, Plass2018_planaria_SingleCellData}. The age of a cell can be interpreted as the distance to the root in the ground-truth hierarchy. Indeed, as illustrated in Figure~\ref{fig:samplesTables} (right), the cells get older towards the boundary. In order to estimate the age of a cell, we compute the heuristic hyperbolic SMT via \mthd{} and measure the tree geodesic distance from any given cell to the root cell. Specifically, we compute the shortest paths over trees via the Dijkstra algorithm \cite{Dijkstra1959}, obtaining a distance vector of dimension $|P|$ storing the distances for each cell -- i.e. their ages estimated via \mthd{}. The latter distance vector is compared to the one obtained from PAGA, via the Euclidean distance between vectors. This defines the distance error used to compare hierarchy discovery methods on the task of cell age prediction.

\begin{figure*}[ht!]
    \centering
  \begin{subfigure}[t]{0.48\linewidth}
     \centering
    \includegraphics[width=1.0\linewidth]{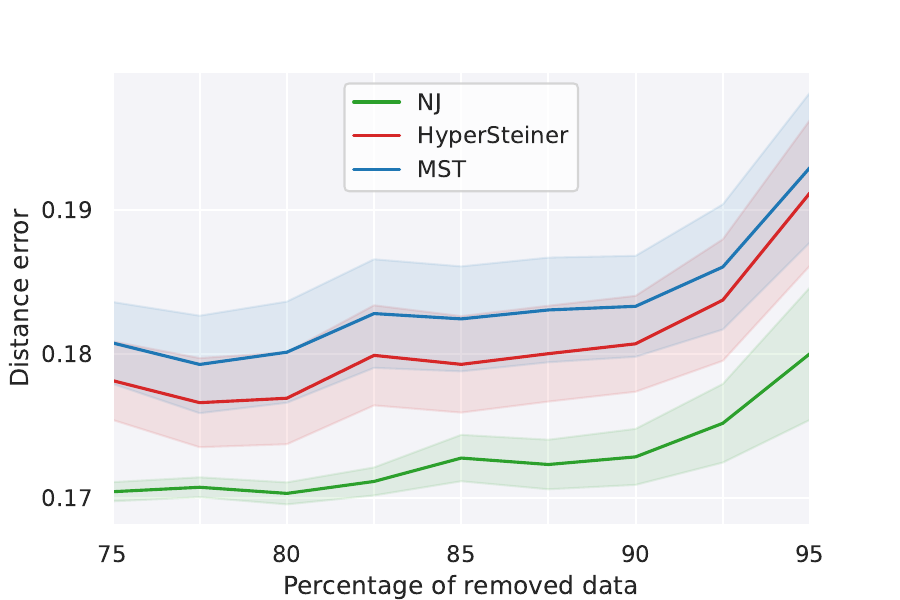}
       \end{subfigure}
  \begin{subfigure}[t]{0.48\linewidth}
    \centering
    \includegraphics[width=1.0\linewidth]{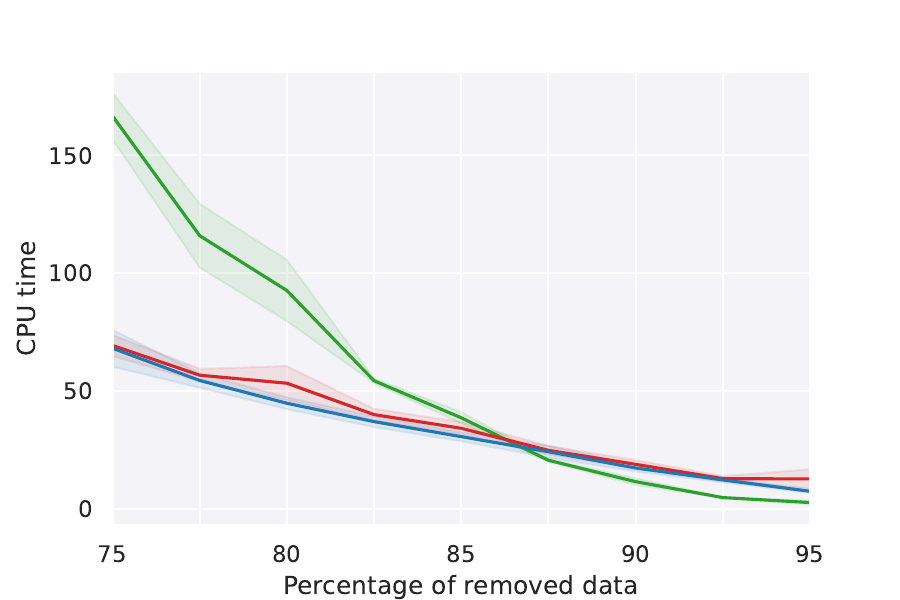}
    \end{subfigure}
     \caption{Comparison for the cell age prediction task for a varying amount of missing data. The plots display mean (lines) over $100$ removals together with standard deviation (shaded regions). \textbf{Left}: error of cell age prediction in terms of Euclidean distance. \textbf{Right}: CPU time in seconds.}
     \label{fig:expplots}
\end{figure*}

We compare \mthd{} to two baselines: the MST and Neighbor Joining (NJ) \cite{Saitou1987neighborjoining} -- a popular distance-based method in computational biology aimed at reconstructing trees from data. We implement the latter via the Biotite Python library \cite{kunzmann2018biotite}. Table~\ref{tab:exptable} reports the results on the whole \emph{Planaria} dataset in terms of the distance error and the CPU time. In order to evaluate the hierarchy discovery capabilities with missing data, we moreover remove random portions of the dataset before inferring the tree. Figure~\ref{fig:expplots} reports results as the portion of removed data varies from 75\% to 95\%. The process is repeated for $100$ random removals per percentage of missing data. Note that Table~\ref{tab:exptable} does not report standard deviations since it involves the whole dataset. 

\begin{table}[!th]
    \centering
    \resizebox{0.8\columnwidth}{!}{%
    \begin{tabular}{lcccc}
        \toprule
        Method &              NJ &     HyperSteiner & MST\\
        \midrule
        Distance error &        0.17  & 0.18 & 0.19\\
        CPU & 12604 & 995 & 914 \\
        \bottomrule
    \end{tabular}%
    }
    \caption{Comparison for the cell age prediction task on the whole \emph{Planaria} dataset.}
    \label{tab:exptable}
\end{table}

\subsubsection{Discussion.}

As evident from Table~\ref{tab:exptable} for the whole dataset and from Figure~\ref{fig:expplots} (left) for missing data, \mthd{} performs better than MST, but NJ outperforms both. However, the runtimes demonstrate that \mthd{} and MST are similar in terms of CPU time, and both scale better than NJ with respect to the amount of data considered. This shows empirically that NJ falls behind \mthd{} in terms of computational complexity. Notably, \mthd{} is more than 12 times faster than NJ on the whole \emph{Planaria} dataset as evident from Table~\ref{tab:exptable}, at the cost of an increase in distance error in the cell age prediction task (0.18 instead of 0.17, i.e., 1.06 times higher). The result is expected from the theory, since the asymptotic time complexity of NJ is $\mathcal{O}(|P|^{3})$ \cite{Studier1988NJ}, while it is $\mathcal{O}(|P| \log |P|)$ for both MST and \mthd{} as discussed in Section~\ref{sec:compcomp}. There exist variations and improvements of NJ -- for example FastNJ \cite{Elias2009FastNJ} and FastTree \cite{Price2009FastTree}. However, they still fall behind \mthd{} in terms of computational complexity, at least asymptotically. Specifically, FastNJ and FastTree have complexity $\mathcal{O}(|P|^2)$ and $\mathcal{O}(|P|^{3/2}\log |P|)$, respectively.

Overall, NJ trades off compute for accuracy, which is a strong practical limitation for large datasets. We believe that the substantial improvement in computational efficiency is a crucial advantage of \mthd{} in practical applications. Trading off performance for scalability with respect to dataset size is often necessary in order to deploy methods from computational geometry on large-scale datasets, and this trade off constitutes a current trend in the field of phylogenetics \cite{Vaz2020DistanceBasedPhylogeneticInference}.

\section{Conclusions and Future Work}
In this work, we have proposed \mthd{} -- an efficient heuristic method for computing Steiner minimal trees in the hyperbolic space by extending the Euclidean SLL algorithm. We have studied the geometric and implementation aspects of \mthd{}. The produced Steiner tree is, by construction, better embedded in the hyperbolic space than the minimum spanning tree. Motivated by this, we have explored applications of \mthd{} to hierarchy discovery in hyperbolic data representations of real-life data. In particular, we have considered Neighbor Joining -- a standard method in phylogenetics -- and shown that, for large datasets, \mthd{} is significantly faster, although at a cost of performance.

As a line for future investigation, we envision applications of \mthd{} to domains beyond phylogenetics. As mentioned in Section \ref{sec:intro}, hierarchies emerge, for example, in natural language processing \cite{nickel2017poincare} and reinforcement learning \cite{cetin2022hyperbolic}. Hyperbolic SMTs can uncover hierarchies and simultaneously infer novel or missing data in the form of Steiner points. Therefore, \mthd{} can potentially provide an efficient tool for analysis and inference. Since these domains often involve large amounts of data, we believe that the scalability of \mthd{} over Neighbor Joining will constitute an essential advantage in practice. 

\section*{Acknowledgements}
We wish to thank Mårten Björkman and Ciwan Ceylan for their feedback. We are also grateful to the anonymous reviewers for their comments. This work has been supported by the Swedish Research Council, the Knut and Alice Wallenberg Foundation, and the European Research Council (ERC AdG BIRD). Alejandro García Castellanos is funded by the Hybrid Intelligence Center, a 10-year programme funded through the research programme Gravitation which is (partly) financed by the Dutch Research Council (NWO).

\bibliographystyle{plain}
\bibliography{references}

\newpage

\onecolumn

\title{\Large Appendix}
\maketitle

\appendix
\section{Mathematical Details}\label{sec:appA}
Here, we provide mathematical details complementing the main body of the work.

\subsection{Isoptic Curves in Klein-Beltrami model.}\label{sec:appisop}
Firstly, we prove Proposition~\ref{prop:isoptic} by deriving the expression for the isoptic curves $\varphi_{x,y,\alpha}$ in the Klein-Beltrami model.  
\begin{proof}
Let $x, y \in \mathbb{K}^2$ and $0<\alpha<\pi$. We can rewrite Equation~\ref{eq:KleinDistance} as:
\begin{equation}\label{eq:coshKleinDistance}
    \operatorname{cosh}\left(d(x,y)\right) = \frac{-\langle x,y\rangle}{\sqrt{\langle x,x \rangle \langle y,y \rangle}}\,.
\end{equation}
Since $\operatorname{sinh}\left(\operatorname{arcosh}(z)\right) = \sqrt{z^2-1}$ for any $z>1$, we obtain:
\begin{equation}\label{eq:sinhKleinDistance}
    \operatorname{sinh}\left(d(x,y)\right) = \sqrt{\left(\frac{\langle x,y\rangle}{\sqrt{\langle x,x \rangle \langle y,y \rangle}}\right)^2-1}\,.
\end{equation}
The desired expression for $\varphi_{x,y,\alpha}$ is obtained after elementary algebraic simplifications by plugging the above equations into the hyperbolic cosine rule:
\begin{align*}
    \operatorname{cosh}\left(d(x,y)\right) &=  \operatorname{cosh}\left(d(x,s)\right) \operatorname{cosh}\left(d(y,s)\right) - \operatorname{sinh}\left(d(x,s)\right)\operatorname{sinh}\left(d(y,s)\right)\operatorname{cos}\left(\alpha\right).
\end{align*}
\end{proof}

Next, we provide the full expression for the quartic $\psi_{x,y, \alpha}$ from Proposition~\ref{prop:quadricklein}, where $x,y \in \mathbb{K}^2$ and $0<\alpha<\pi$. The expression, which is obtained by reorganizing Equation~\ref{eq:system} and squaring both hand sides, is:  
\begin{align*}
\psi_{x,y, \alpha} (s) &= \langle x, s \rangle^2 \langle y, s \rangle^2 + \langle x, y \rangle^2 \langle s, s \rangle^2  - 2 \langle x, s \rangle \langle y, s \rangle \langle x, y \rangle \langle s, s \rangle -\\
&- \cos(\alpha)^2 \langle x, s \rangle^2 \langle y, s \rangle^2  - \cos(\alpha)^2 \langle x, x \rangle \langle y, y \rangle \langle s, s \rangle^2   + \cos(\alpha)^2 \langle y, s \rangle^2 \langle x, x \rangle \langle s, s \rangle   +  \cos(\alpha)^2 \langle x, s \rangle^2 \langle y, y \rangle \langle s, s \rangle . 
\end{align*}

Lastly, we can check in constant time if there exists a Steiner point  for any hyperbolic triangle $\triangle p_1 p_2 p_3$ by verifying that all the inner angles $(\theta_1, \theta_2, \theta_3)$ are less than $2 \pi / 3$ using the hyperbolic cosine rule:
\[
\theta_i = \arccos\left(\frac{\operatorname{cosh}\left(d(p_i, p_j)\right)\operatorname{cosh}\left(d(p_i, p_k)\right) - \cosh\left(d(p_j, p_k)\right)}{\operatorname{sinh}\left(d(p_i, p_j)\right)\operatorname{sinh}\left(d(p_i, p_k)\right)}\right)\,.
\]

\subsection{Isoptic Curves in Other Hyperbolic Models.}\label{sec:appother}

As we can see, the previous proof mainly relies on the hyperbolic cosine rule; hence, we can generalize the previous steps in order to obtain the isoptic curves for any hyperbolic model. As an example, we will see how we can obtain the isoptic curve for the upper half-plane model $\HH^2= \{x+iy \in \CC \mid y>0\}$.

Firstly, recall that in the upper half-plane model, we can express the distance between two points explicitly as follows, with $\overline{p}$ denoting the complex conjugate of $p$:
\begin{prop}[{\cite{xu_introduction_2021}}]
For any points $p=x_1 + iy_1$ and $q= x_2 + iy_2$ in $\HH^2$, their hyperbolic distance is given by the following formula:
\begin{align*}
d_{\HH}(p, q) &= \abs{\log \frac{\abs{\overline{p} - q} + \abs{p -q}}{\abs{\overline{p} - q} - \abs{p - q}}}\\
&=\abs{\log \frac{\sqrt{(x_2 -x_1)^2 + (y_2 + y_1)^2} + \sqrt{(x_2 -x_1)^2 + (y_2 - y_1)^2}}{\sqrt{(x_2 -x_1)^2 + (y_2 + y_1)^2} - \sqrt{(x_2 -x_1)^2 + (y_2 - y_1)^2}}}\,.
\end{align*}
\end{prop}

Hence, we also obtain the following formulas:
\begin{align}
\operatorname{cosh}\left(d_{\HH}(p, q)\right) = \frac{(x_1 -x_2)^2 + y_1^2 + y_2^2}{2 y_1 y_2}\,, \label{eq:coshHypUH}\\[10pt]
\operatorname{sinh}\left(d_{\HH}(p, q)\right) = \frac{\sqrt{(x_2 -x_1)^2 + (y_2 - y_1)^2} \  \sqrt{(x_2 -x_1)^2 + (y_2 + y_1)^2}}{2 y_1 y_2}\,. \label{eq:sinhHypUH}    
\end{align}

Substituting \ref{eq:coshHypUH}, \ref{eq:sinhHypUH} into the hyperbolic cosine rule and simplifying, we obtain the following characterization of the isoptic curves in the upper half-plane model.

\begin{prop}\label{prop:isopticUpper}
Given two points $p, q \in \mathcal{X}=\HH^2$ and an angle $0<\alpha<\pi$, the isoptic curve $C_\alpha(p,q)$ in the upper-half model is given by $\{s=x+iy\in \HH^2 \mid \tilde{\varphi}_{p,q, \alpha}(s) = 0\}$, where: 
 \begin{align*}
  \tilde{\varphi}_{p,q, \alpha}(s) &=   ((x_1 -x)^2 + y_1^2 + y^2)((x_2 -x)^2 + y_2^2 + y^2) -2y^2((x_1 - x_2)^2 + y_1^2 + y_2^2)\\
     &-\cos(\alpha)\sqrt{((x_1 -x)^2 + (y_1 -y)^2)((x_1 -x)^2 + (y_1 + y)^2)((x_2 -x)^2 + (y_2 -y)^2)((x_2 -x)^2 + (y_2 + y)^2)}
\end{align*}
\end{prop}

Similarly to the Klein-Beltrami model, we can remove the square root from the equation $\tilde{\varphi}_{p,q, 2\pi/3} = 0$ by rearranging the terms and squaring. The level set of the resulting degree $8$ polynomial $\tilde{\psi}_{p,q, 2\pi/3}$ is equivalent to the one obtained by \cite{halverson2005steiner}. As stated by the latter, this level set corresponds to the locus of points that form angles of either $2 \pi / 3$ or $\pi / 3$ with $p$ and $q$, i.e., the union of isoptic curves $C_{2\pi/3}(p,q) \cup C_{\pi/3}(p,q)$.

\subsection{Discussion of Hyperbolic Models.}\label{sec:klein-upper}

In our exploration, we opted for the Klein-Beltrami model over the upper half-plane model suggested by \cite{halverson2005steiner}. Several methodological considerations influenced this decision:

\begin{itemize}
    \item \begin{sloppypar}
     Examination of Figure~\ref{fig:curvesHalf} reveals that the isoptic curves associated with the upper half-plane model exhibit greater complexity compared to those derived from the Klein-Beltrami model. Precisely, the polynomial $\tilde{\psi}_{p,q, 2\pi/3}\in \RR[x,y]$ in the upper half-plane model attains a degree of eight, while its counterpart ${\psi_{p,q, 2\pi/3}\in \RR[x,y]}$ in the Klein-Beltrami model is of degree four. Furthermore, the upper half-plane model necessitates restricting the level curves to $\{(x,y) \in \RR^2 \mid y >0\}$, while the Klein-Beltrami model naturally contains isoptic curves within the unit open ball. Consequently, solving the associated system of equations in the upper half-plane model requires constrained optimization solvers to avoid spurious solutions.
   
    \end{sloppypar}

\begin{figure}[ht!]
\centering
  \begin{subfigure}[t]{0.5\textwidth}
     \centering
     \includegraphics[width=\linewidth]{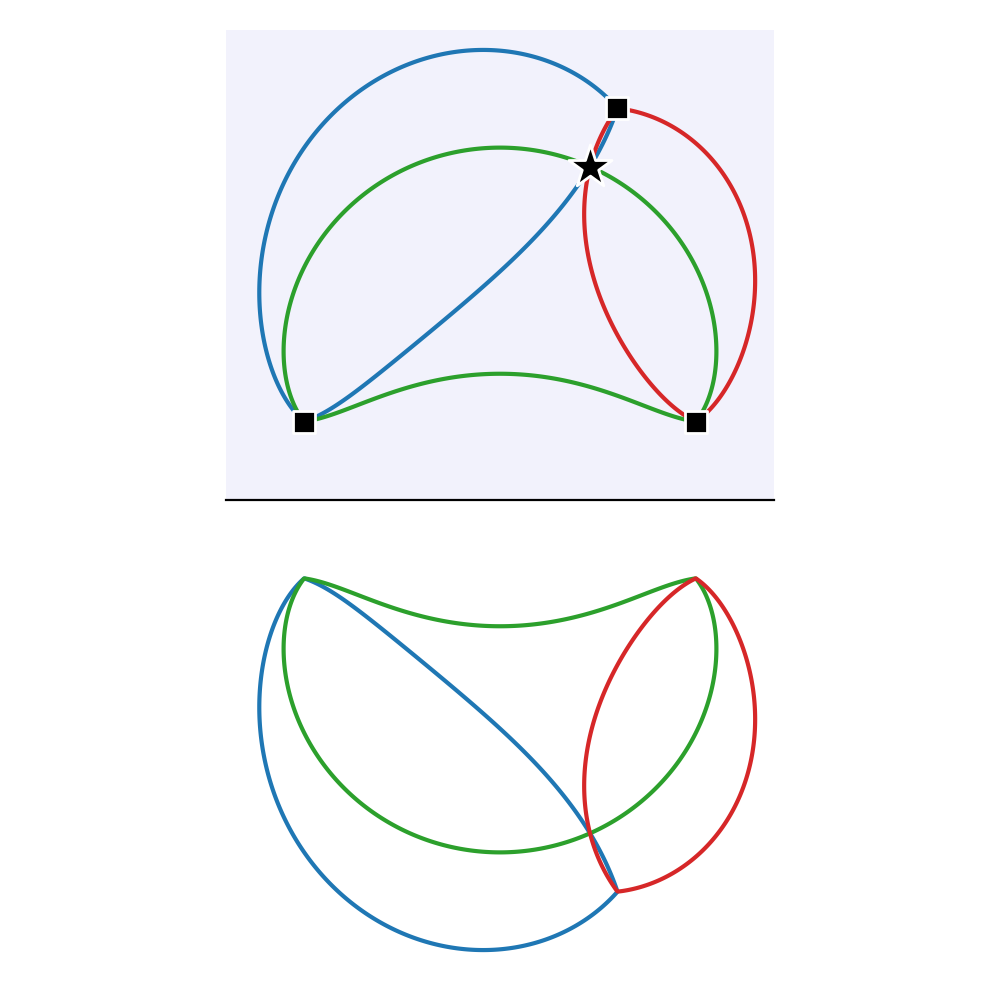}
     \caption*{$\left\{s\in \CC \mid \tilde{\varphi}_{p,q,\frac{2\pi}{3}}(s) = 0\right\}$}
     \label{fig:isopticsHalf}
 \end{subfigure}\hfill
 \begin{subfigure}[t]{0.5\textwidth}
     \centering
     \includegraphics[width=0.57\textwidth]{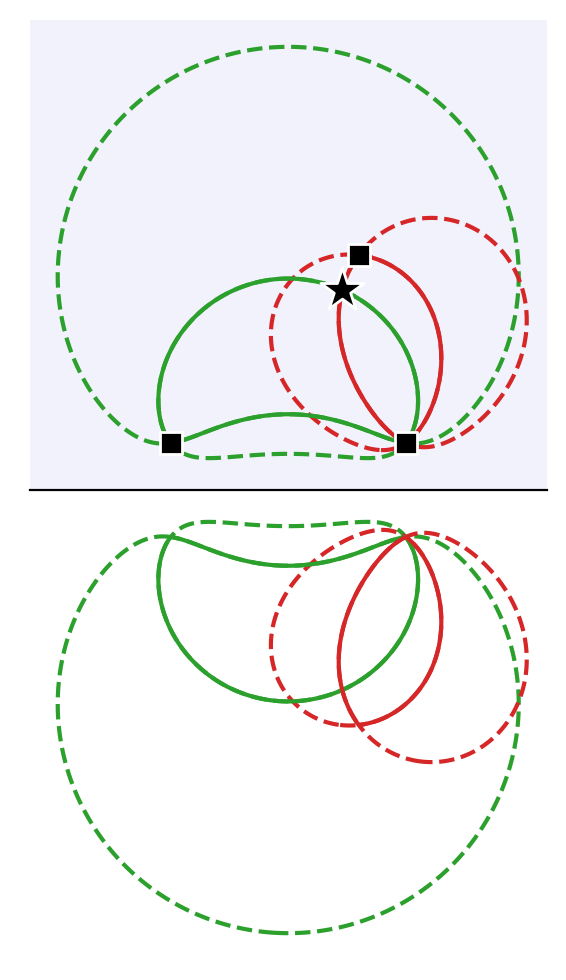}
     \caption*{$\left\{s\in \CC \mid \tilde{\psi}_{p,q,\frac{2\pi}{3}}(s) = 0\right\}$}
     \label{fig:polyHalf}
 \end{subfigure}
\caption{Example of isoptic curves in the upper half-plane model $\HH^2$. Square ($\,\blacksquare\,$) denotes terminals, star ($\bigstar$) denotes Steiner points, the solid line corresponds to the isoptic curve for $\alpha=2\pi/3$, and the dashed line to the one for the complementary angle. The shaded area corresponds to $\HH^2$.}
\label{fig:curvesHalf}
\end{figure}

    \item As discussed in Section~\ref{sec:hyperdel}, the computation of the Delaunay triangulation is viable in the Klein-Beltrami model using a computationally efficient method outlined in \cite{HyperbolicVoronoiDiagramsMadeEasy2010}. Therefore, this approach is more coherent since both the Delaunay triangulation and the isoptic curves are computed within the same model. This eliminates the need to map points between models, which can be computationally expensive with a large number of samples.
\end{itemize}

\section{Additional Experimental Results}

\subsection{Scalability Analysis on Synthetic Data.}

In this section, we further expand the results obtained in Section~\ref{sec:scalability} by considering additional measures. Specifically, we look at both the number of 3- and 4-point local solutions used in Step~5 of the SLL algorithm (Algorithm \ref{alg:sll-alg}). \\

Firstly, Table~\ref{tab:scalability_analysisCenteredFULL} shows that when we compute 4-point FSTs in \mthd{}, the number of local solutions is similar to the Euclidean case. Secondly, the number of additional FSTs when deploying the homotopy-continuation method is minor (see Table~\ref{tab:scalability_analysisCenteredFULL} and Table~\ref{tab:scalability_analysisIdealFULL}). This reinforces our argument that opting for the setup 4-Simple instead of 4-Precise does not lead to a significant performance loss.

\begin{table}[!ht]
\centering
\resizebox{\textwidth}{!}{
\begin{tabular}{lcccccccccccccccccc}
\toprule
\multicolumn{5}{c}{} & \multicolumn{14}{c}{Klein-Beltrami}  \\
\cmidrule(lr){6-19} 
\multicolumn{1}{c}{} & \multicolumn{4}{c}{Euclidean} & \multicolumn{3}{c}{3+Simple} & \multicolumn{3}{c}{3+Precise} & \multicolumn{4}{c}{4+Simple} & \multicolumn{4}{c}{4+Precise} \\
\cmidrule(lr){2-5} \cmidrule(lr){6-8} \cmidrule(lr){9-11} \cmidrule(lr){12-15} \cmidrule(lr){16-19}  
$|P|$ &   \# 3 pts &  \# 4 pts &              RED &      CPU &   \# 3 pts &              RED &      CPU &   \# 3 pts &              RED &      CPU &   \# 3 pts &  \# 4 pts &              RED &      CPU &   \# 3 pts &  \# 4 pts &              RED &       CPU \\
\midrule
50    &    $12.29$ &    $1.50$ &  $2.47 \pm 0.55$ &   $0.03$ &    $13.98$ &  $2.25 \pm 0.61$ &   $0.05$ &    $14.19$ &  $2.31 \pm 0.61$ &   $0.18$ &    $12.21$ &    $1.45$ &  $2.48 \pm 0.61$ &   $0.21$ &    $12.38$ &    $1.50$ &  $2.55 \pm 0.60$ &    $1.29$ \\
100   &    $25.15$ &    $2.86$ &  $2.46 \pm 0.40$ &   $0.06$ &    $28.19$ &  $2.17 \pm 0.42$ &   $0.08$ &    $28.54$ &  $2.21 \pm 0.43$ &   $0.30$ &    $24.70$ &    $2.79$ &  $2.39 \pm 0.44$ &   $0.45$ &    $24.90$ &    $2.89$ &  $2.45 \pm 0.45$ &    $2.49$ \\
500   &   $126.54$ &   $15.39$ &  $2.60 \pm 0.22$ &   $0.37$ &   $142.88$ &  $2.27 \pm 0.21$ &   $0.42$ &   $145.33$ &  $2.33 \pm 0.20$ &   $1.71$ &   $123.99$ &   $15.36$ &  $2.51 \pm 0.20$ &   $2.28$ &   $125.56$ &   $15.94$ &  $2.58 \pm 0.20$ &   $13.07$ \\
1,000  &   $254.62$ &   $30.86$ &  $2.62 \pm 0.17$ &   $0.83$ &   $288.57$ &  $2.32 \pm 0.16$ &   $1.08$ &   $293.13$ &  $2.37 \pm 0.16$ &   $3.56$ &   $249.10$ &   $31.98$ &  $2.56 \pm 0.17$ &   $5.00$ &   $251.80$ &   $33.19$ &  $2.62 \pm 0.17$ &   $26.80$ \\
5,000  &  $1278.59$ &  $155.37$ &  $2.70 \pm 0.07$ &   $9.00$ &  $1446.33$ &  $2.36 \pm 0.07$ &   $7.86$ &  $1466.95$ &  $2.43 \pm 0.07$ &  $19.04$ &  $1252.59$ &  $155.92$ &  $2.61 \pm 0.07$ &  $29.62$ &  $1264.92$ &  $161.90$ &  $2.68 \pm 0.07$ &  $137.45$ \\
10,000 &  $2559.78$ &  $308.65$ &  $2.73 \pm 0.06$ &  $29.84$ &  $2894.78$ &  $2.39 \pm 0.06$ &  $23.15$ &  $2930.69$ &  $2.45 \pm 0.06$ &  $45.52$ &  $2506.24$ &  $313.90$ &  $2.64 \pm 0.06$ &  $74.05$ &  $2526.72$ &  $324.59$ &  $2.70 \pm 0.06$ &  $291.53$ \\
\midrule
 Avg.  & & & $2.60 \pm 0.32$ &   & & $2.29 \pm 0.33$ &    & & $2.35 \pm 0.33$ &   & & & $2.53 \pm 0.34$ & & & & $2.60 \pm 0.34$ & \\
\bottomrule
\end{tabular}
}
\caption{Scalability analysis of centered Gaussians ($\mathcal{N}(0, 0.5)$ for Euclidean and $\mathcal{G}(0, 0.5)$ for Hyperbolic). \#3: number of 3-point local solutions used. \#4: number of 4-point local solutions used. RED: reduction over MST (\%). CPU: total CPU time (sec.). }
\label{tab:scalability_analysisCenteredFULL}
\end{table}

\begin{table}[!ht]
\centering
\resizebox{\textwidth}{!}{
\begin{tabular}{lcccccccccccccc}
\toprule
{} & \multicolumn{3}{c}{3+Simple} & \multicolumn{3}{c}{3+Precise} & \multicolumn{4}{c}{4+Simple} & \multicolumn{4}{c}{4+Precise} \\
 \cmidrule(lr){2-4} 
 \cmidrule(lr){5-7} 
 \cmidrule(lr){8-11} 
 \cmidrule(lr){12-15} 
 $|P|$ & \# 3 pts & RED & CPU & \# 3 pts & RED &  CPU & \# 3 pts & \# 4 pts & RED & CPU & \# 3 pts & \# 4 pts & RED & CPU \\
\midrule
50    &    $12.89$ &  $2.61 \pm 0.67$ &   $0.03$ &    $13.29$ &  $2.75 \pm 0.67$ &   $0.21$ &    $10.58$ &    $1.89$ &  $3.00 \pm 0.73$ &   $0.21$ &    $10.79$ &    $2.04$ &  $3.18 \pm 0.72$ &    $2.47$ \\
100   &    $25.99$ &  $2.56 \pm 0.49$ &   $0.07$ &    $26.80$ &  $2.68 \pm 0.48$ &   $0.41$ &    $22.05$ &    $3.15$ &  $2.85 \pm 0.51$ &   $0.46$ &    $22.62$ &    $3.27$ &  $2.99 \pm 0.49$ &    $4.21$ \\
500   &   $140.88$ &  $2.30 \pm 0.18$ &   $0.42$ &   $143.81$ &  $2.39 \pm 0.19$ &   $2.05$ &   $122.54$ &   $14.90$ &  $2.54 \pm 0.20$ &   $2.36$ &   $124.74$ &   $15.41$ &  $2.63 \pm 0.21$ &   $18.22$ \\
1,000  &   $281.17$ &  $2.29 \pm 0.16$ &   $1.06$ &   $286.07$ &  $2.37 \pm 0.16$ &   $4.36$ &   $243.09$ &   $30.94$ &  $2.53 \pm 0.16$ &   $5.07$ &   $246.72$ &   $31.49$ &  $2.61 \pm 0.15$ &   $35.16$ \\
5,000  &  $1427.31$ &  $2.34 \pm 0.07$ &   $7.80$ &  $1441.87$ &  $2.39 \pm 0.08$ &  $24.38$ &  $1241.00$ &  $152.46$ &  $2.57 \pm 0.07$ &  $30.45$ &  $1252.94$ &  $153.44$ &  $2.62 \pm 0.07$ &  $172.60$ \\
10,000 &  $2865.34$ &  $2.36 \pm 0.05$ &  $22.66$ &  $2883.35$ &  $2.39 \pm 0.04$ &  $56.83$ &  $2491.04$ &  $304.27$ &  $2.59 \pm 0.05$ &  $71.84$ &  $2505.13$ &  $306.48$ &  $2.63 \pm 0.05$ &  $345.08$ \\
\midrule
 Avg.  & & $2.41 \pm 0.37$ &   & & $2.48 \pm 0.38$ &    & & & $2.68 \pm 0.42$ &   & & & $2.78 \pm 0.43$ & \\
\bottomrule
\end{tabular}
}
\caption{Scalability analysis of mixture of $\mathcal{G}(\mu_{15, k}(0.9), 0.5)$, $k \in \{1, \cdots, 15\}$. \#3: number of 3-point local solutions used. \#4: number of 4-point local solutions used. RED: reduction over MST (\%). CPU: total CPU time (sec.). }
\label{tab:scalability_analysisIdealFULL}
\end{table}

\newpage

As an additional experiment, we evaluate \mthd{} on a synthetic dataset generated from rejection sampling of a uniform distribution $U([-1,1]^2)$ on the open unit ball (see Table~\ref{tab:scalability_analysisUniFULL}).

\begin{table*}[!ht]
\centering
\resizebox{\textwidth}{!}{
\begin{tabular}{lcccccccccccccc}
\toprule
{} & \multicolumn{3}{c}{3+Simple} & \multicolumn{3}{c}{3+Precise} & \multicolumn{4}{c}{4+Simple} & \multicolumn{4}{c}{4+Precise} \\
 \cmidrule(lr){2-4} 
 \cmidrule(lr){5-7} 
 \cmidrule(lr){8-11} 
 \cmidrule(lr){12-15} 
 $n$ & \# 3 pts & RED & CPU & \# 3 pts & RED &  CPU & \# 3 pts & \# 4 pts & RED & CPU & \# 3 pts & \# 4 pts & RED & CPU \\
\midrule
50    &    $14.07$ &  $2.08 \pm 0.66$ &   $0.04$ &    $14.33$ &  $2.15 \pm 0.67$ &   $0.15$ &    $12.30$ &    $1.45$ &  $2.32 \pm 0.75$ &   $0.21$ &    $12.46$ &    $1.50$ &  $2.39 \pm 0.76$ &    $1.55$ \\
100   &    $27.85$ &  $2.06 \pm 0.43$ &   $0.07$ &    $28.41$ &  $2.14 \pm 0.45$ &   $0.33$ &    $24.43$ &    $2.76$ &  $2.28 \pm 0.46$ &   $0.45$ &    $24.57$ &    $3.07$ &  $2.38 \pm 0.47$ &    $3.01$ \\
500   &   $142.36$ &  $2.12 \pm 0.25$ &   $0.45$ &   $145.18$ &  $2.20 \pm 0.24$ &   $1.78$ &   $123.90$ &   $15.07$ &  $2.34 \pm 0.26$ &   $2.52$ &   $125.69$ &   $15.71$ &  $2.42 \pm 0.25$ &   $14.76$ \\
1,000  &   $285.19$ &  $2.14 \pm 0.20$ &   $0.92$ &   $290.41$ &  $2.21 \pm 0.20$ &   $3.61$ &   $247.77$ &   $30.38$ &  $2.35 \pm 0.21$ &   $4.96$ &   $251.03$ &   $31.79$ &  $2.44 \pm 0.22$ &   $29.19$ \\
5,000  &  $1437.89$ &  $2.22 \pm 0.09$ &   $7.54$ &  $1459.21$ &  $2.28 \pm 0.09$ &  $20.92$ &  $1246.01$ &  $155.36$ &  $2.44 \pm 0.10$ &  $30.16$ &  $1260.73$ &  $159.93$ &  $2.51 \pm 0.10$ &  $150.44$ \\
10,000 &  $2888.47$ &  $2.25 \pm 0.07$ &  $22.08$ &  $2923.18$ &  $2.30 \pm 0.07$ &  $48.44$ &  $2499.72$ &  $312.91$ &  $2.48 \pm 0.07$ &  $71.81$ &  $2522.73$ &  $320.71$ &  $2.53 \pm 0.07$ &  $306.32$ \\
\midrule
 Avg.  & & $2.14 \pm 0.35$ &   & & $2.21 \pm 0.36$ &    & & & $2.37 \pm 0.39$ &   & & & $2.44 \pm 0.39$ & \\
\bottomrule
\end{tabular}
}
\caption{Scalability analysis of rejection sampling of a uniform distribution $U([-1,1]^2)$ on the open unit ball. \#3: number of 3-point local solutions used. \#4: number of 4-point local solutions used. RED: reduction over MST (\%). CPU: total CPU time (sec.). }
\label{tab:scalability_analysisUniFULL}
\end{table*}

\end{document}